 \documentclass[twocolumn,showpacs,superscriptaddress,aps,prc,10pt,nofootinbib]{revtex4-1} 
 \usepackage[pdftex]{graphicx} 
 \usepackage{cancel} 
 \usepackage[usenames,svgnames,table]{xcolor} 
 \usepackage[english]{babel} 
 \usepackage{ucs} 
 \usepackage[utf8]{inputenc} 
 \usepackage{siunitx} 
 \usepackage[version=3]{mhchem} 
 \usepackage{multirow} 
 \usepackage{textcomp} 
 \usepackage{soul} 
 \usepackage{dcolumn}% Align table columns on decimal point 
 \usepackage{bm}% bold math 
 \usepackage{mciteplus} %multiple citation with same ref number 

\newcommand{\gar}{\textsc{Garfield}} 
\newcommand{\rc}{\textsc{RCo}} 
\newcommand{\gemini}{\textsc{Gemini++}} 
\newcommand{\hf}{\textsc{HF$\ell$}} 
\newcommand{\al}{$\alpha$} 
 
\newcommand{\er}{$ER$} 
\newcommand{\est}{$E^\ast$} 
\newcommand{\cb}{$^{13}$C} 
\newcommand{\ca}{$^{12}$C} 
\newcommand{\mgb}{$^{25}$Mg} 
\newcommand{\mga}{$^{24}$Mg} 
 
\begin{document} 
%  
% \preprint{APS/123-QED} 
%  
\title{Study of well selected evaporation chains in the decay of excited \mgb{}}

\author{A.~Camaiani}  \email{alberto.camaiani@fi.infn.it} 
  \affiliation{Dipartimento di Fisica,  Universit\`a di Firenze, Italy} 
  \affiliation{INFN, Sezione di Firenze, Italy} 
\author{G.~Casini} 
  \affiliation{INFN, Sezione di Firenze, Italy} 
\author{L.~Morelli} 
  \affiliation{Dipartimento di Fisica,  Universit\`a di Bologna, Italy} 
  \affiliation{INFN, Sezione di Bologna, Italy} 
  \author{S.~Barlini} 
  \affiliation{Dipartimento di Fisica,  Universit\`a di Firenze, Italy} 
  \affiliation{INFN, Sezione di Firenze, Italy} 
 \author{S.~Piantelli} 
 \affiliation{INFN, Sezione di Firenze, Italy} 
 \author{G.~Baiocco}
 \affiliation{Dipartimento di Fisica,  Universit\`a di Pavia, Italy} 
 \affiliation{INFN, Sezione di Pavia, Italy}  
\author{M.~Bini} 
  \affiliation{Dipartimento di Fisica,  Universit\`a di Firenze, Italy} 
  \affiliation{INFN, Sezione di Firenze, Italy} 
\author{M.~Bruno} 
\affiliation{Dipartimento di Fisica,  Universit\`a di Bologna, Italy} 
  \affiliation{INFN, Sezione di Bologna, Italy} 
   \author{A.~Buccola} 
  \affiliation{Dipartimento di Fisica,  Universit\`a di Firenze, Italy} 
  \affiliation{INFN, Sezione di Firenze, Italy} 
\author{M.~Cinausero} 
 \affiliation{INFN, Laboratori Nazionali di Legnaro, Italy} 
\author{M.~Cicerchia} 
 \affiliation{INFN, Laboratori Nazionali di Legnaro, Italy} 
\author{M.~D'Agostino} 
  \affiliation{Dipartimento di Fisica,  Universit\`a di Bologna, Italy} 
  \affiliation{INFN, Sezione di Bologna, Italy}  
\author{M.~Degelier} 
  \affiliation{Physics Departement, Univeristy of Nevsehir, Science and Art Faculty, Nevsehir, Turkey} 
\author{D.~Fabris} 
\affiliation{INFN, Sezione di Padova, Italy} 
\author{C.~Frosin} 
    \affiliation{Dipartimento di Fisica,  Universit\`a di Firenze, Italy} 
  \affiliation{INFN, Sezione di Firenze, Italy} 
% \author{N.~Gelli}% 
%   \affiliation{INFN, Sezione di Firenze, Italy} 
\author{F.~Gramegna} 
 \affiliation{INFN, Laboratori Nazionali di Legnaro, Italy}   
 \author{F.~Gulminelli}
 \affiliation{LPC (IN2P3-CNRS/Ensicaen et Universit\'e), F-14076 Caen c\'edex, France}
% \author{I.~Lombardo} 
%  \affiliation{INFN, Sezione di Catania, Italy} 
 \author{G.~Mantovani}
  \affiliation{INFN, Laboratori Nazionali di Legnaro, Italy} 
\author{T.~Marchi} 
 \affiliation{INFN, Laboratori Nazionali di Legnaro, Italy} 
\author{A.~Olmi}% 
\affiliation{INFN, Sezione di Firenze, Italy} 
\author{P.~Ottanelli} 
  \affiliation{Dipartimento di Fisica,  Universit\`a di Firenze, Italy} 
  \affiliation{INFN, Sezione di Firenze, Italy} 
\author{G.~Pasquali} 
  \affiliation{Dipartimento di Fisica,  Universit\`a di Firenze, Italy} 
  \affiliation{INFN, Sezione di Firenze, Italy} 
\author{G.~Pastore} 
  \affiliation{Dipartimento di Fisica,  Universit\`a di Firenze, Italy} 
  \affiliation{INFN, Sezione di Firenze, Italy} 
\author{S.~Valdr\'e} 
\affiliation{INFN, Sezione di Firenze, Italy}
\author{G.~Verde} 
 \affiliation{INFN, Sezione di Catania, Italy} 
 
\begin{abstract} 
The reaction $^{12}$C+$^{13}$C at 95$\,$MeV bombarding energy is
studied using the GARFIELD~+~Ring~Counter apparatus located at the INFN 
Laboratori Nazionali di    
Legnaro. 
In this paper we want to investigate the de-excitation  of \mgb{}  %$^{25}$Mg 
aiming both at a new stringent test of the statistical description 
of nuclear decay and a direct comparison with the decay of the system \mga{} 
formed through \ca{}+\ca{} reactions previously studied. 
Thanks to the large acceptance of the 
detector and to its good fragment identification capabilities, we could apply 
stringent selections on fusion-evaporation events, requiring their completeness in charge. 
The main decay features of the evaporation residues and of
the emitted light particles 
are overall well described by a pure statistical model; however, as for 
the case of the previously studied  \mga{}, we  
observed some  deviations in the branching ratios, in particular for those 
chains involving only the evaporation of \al{} particles. 
From this point of view the 
behaviour  of the $^{24}$Mg and $^{25}$Mg decay cases  
appear to be rather similar. An attempt to obtain a full mass 
balance even without neutron detection is also discussed. 
\end{abstract}

\maketitle

\section{Introduction}

Reactions with light nuclei have been extensively studied during the past,
even at  bombarding energies below 100$\,$MeV~\cite{bib:heusch81,bib:von11}, for several
reasons. From the technical point of view 
events with limited number of light fragments are easier 
to detect and characterize. 
On the theoretical side, it is interesting to
verify the  
applicability of statistical concepts to the decay of systems formed 
by a moderate number of nucleons. Moreover, light nuclei,  
especially those with $N$=$Z$, manifest in their low-lying structure 
evident clusterization effects which can still persist, but more loosely, 
with increasing excitation. Therefore, efforts have been done both
theoretically to describe the nature of these quantum systems (in terms
 of clusters~\cite{bib:funaki17})
and experimentally to find signatures of the clusterization
effects also at relatively high excitations~\cite{bib:morelli1, bib:morelli2,bib:vadas15,
  bib:manna16}.   
In recent years, the interest on this subject of nuclear physics
has been renewed~\cite{bib:beck14} thanks to progressively more
sophisticated model approaches 
and to more comprehensive experiments, aiming at the complete
detection and identification of the various ejecta emerging from the
collisions. This  allows to finely select and characterize 
the excited light nuclei; their characteristics can be studied 
not only on average, for rough classes of events, but also in a very
exclusive way following, for instance, the various decay paths for
compound nuclei formed in fusion-like reactions. In this respect, the
analysis of correlations among the detected particles is quite illuminating
because one can try to reconstruct the  intermediate nuclei and their
states populated during the de-excitation.  

In  a more general context, fragment spectroscopy and the
particle correlations after  nuclear collisions represent
powerful tools to study the
decay of transient systems possibly formed embedded in a nuclear
environment. Whether and how the properties of
nuclear resonances and states  
are modified when fragments are formed and disrupted within 
a nuclear medium is not well known and the most typical example
concerns the Hoyle states in autoconjugate nuclei. Indeed, various efforts
have been done 
or are in progress in this direction using reactions at
various bombarding regimes,  
from Tandem energies like the one in this paper~\cite{bib:more16, bib:more_conf16} 
to higher energies~\cite{bib:radu11, bib:more_conf16, bib:quattro_iwm, bib:kaur17, 
  bib:borde16,bib:schmidth17,bib:aquila17}.
  
In this work we give a further contribution to the experimental study
of light excited 
systems, following the approach of our previous papers 
~\cite{bib:morelli1, bib:morelli2,  bib:more_conf14, bib:more_conf16}.
Typically, the method consists of investigating   the various decay
paths of compound nuclei formed in fusion 
reactions, selected as accurately as possible, in order to evidence 
deviations from  the prediction of pure statistical models,  which are
based on    average phase-space considerations and do not include, in their 
``standard''    implementations, the possibility that the branching ratio 
towards $\alpha$ emission might be affected by a possible $\alpha$-structure of the parent state.
The main focus in our previous works was on the fusion reactions 
$^{12}$C+$^{12}$C, producing \mga{} nuclei at \est{}=61.4$\,$MeV excitation 
energy~\cite{bib:morelli1, bib:morelli2}.    
Although we there verified a quite nice  agreement of many decay features 
with the  
predictions of  a statistical model, we observed some 
deviations in the channels involving the evaporation of only $\alpha$ 
particles.  Similar behaviour was found for the same
compound system 
but formed through the reaction $^{14}$N + $^{10}$B at the same excitation 
energy. Such additional finding suggests that the deviations are not (or only 
partially) due 
to the $N$=$Z$ symmetry  of the entrance channel, but pertain to the $N$=$Z$ compound 
nucleus itself~\cite{bib:more_conf14, bib:more_conf16}.    

Therefore, it was quite straightforward  to extend our exploration  to
neighbouring  systems produced at  
comparable excitations again through fusion reactions, but with an 
additional neutron which breaks the $N$=$Z$  symmetry. 
So this paper reports on  the fusion 
reaction $^{12}$C+$^{13}$C at 95$\,$MeV   
bombarding energy, forming  \mgb{} nuclei at 
E$^{*}$=65.7$\,$MeV.  
Specifically, we investigate  how the additional neutron  affects the decay chains of 
\mgb{} compound nuclei. This is done by comparing the new results with both the predictions of the statistical model 
and with the previous experimental results on \mga{} by our apparatus. 
The main result of the paper is that the same anomaly observed for \mga{}
persists for \mgb{}. This could be tentatively understood from the fact that 
the neutron emission is the 
most exothermic decay for the excited \mgb{} (Q$_{val}$=+8.98$\,$MeV)
besides the gamma emission (not seen by our apparatus), thus 
the probability to populate an excited \mga{} starting from a \mgb{} compound is high.
However, as we will show in detail, such a first chance neutron 
emission is far from being the dominant decay channel. 
Rather, the decay pattern of \mgb{} is shown to closely follow the one from \mga{},
with the extra neutron being preferentially emitted from the (neutron rich) evaporation residue. 
In this paper we also present a new analysis technique with the aim of 
estimating the number of evaporated free neutrons 
which are not detected by our apparatus. 
% The idea is to progressively
% build up an experimental system (detectors and analysis tools)
% allowing to precisely select and study the evaporation chains in order
% to perform stringent tests models and to unravel interesting details on
% clusters and correlations within the nuclei.

The paper   is organized  as follows. In Section~\ref{sec:exp} the 
characteristics of the experimental   
apparatus are briefly summarized. 
Section~\ref{sec:sel_and_mc} describes
the criteria adopted in the analysis 
to select   the experimental sample of fusion events and  briefly reminds the 
models used to describe the fusion-evaporation events;
Section~\ref{sec:12cpol}   discusses the problem of the background 
of \ca{}   reactions in  the data on \cb{}. 
The general features of the selected fusion events are presented in 
Section~\ref{sec:fus} where they are compared with  the statistical code predictions. 
The detailed a\-na\-ly\-sis of the various decay chains, the 
comparison between the data of \mgb{} and \mga{} and the main findings  are 
discussed in Section~\ref{sec:dev}, ~\ref{sec:confr} and \ref{sec:jacobi}. A summary of the work is  given  
in Section~\ref{sec:concl}. 
 
\section{\label{sec:exp}Experimental Apparatus} 
 
The experimental apparatus consists of the multi-detector \gar{} and 
Ring Counter, located in the Hall III of Laboratori Nazionali di Legnaro (LNL): 
a complete description and details can be found 
elsewhere~\cite{bib:bruno}. Briefly, \gar{} is a two-stage detector consisting of two 
identical micro-strip gas chambers   
(the $\Delta E$ stage)  and 
CsI(Tl) crystals (for residual particle energy). The chambers allow   
particle identification through $\Delta$E-E correlations and Pulse 
Shape Analysis (PSA) in CsI(Tl).  
The Ring Counter  (hereafter referred to as \rc{}) is a three-stage hodoscope 
fully equipped with digital electronics: it is made of a 8-sector Ionization 
Chamber (IC), followed by  segmented   
reverse mounted silicon detectors and, as last stage, CsI(Tl) crystals. 
The \rc{} covers the polar range from 7$\,^{\circ}$ up to 17$\,^{\circ}$ 
while the \gar{}  geometry covers the angular range from 30$\,^{\circ}$ to 150$\,^{\circ}$ 
% (with 
% a dead zone for target mounting from 83$\,^{\circ}$  to 97$\,^{\circ}$) 
with 180 CsI(Tl) crystals. 
Both detectors have a complete azimuthal symmetry. The  
combination of the two devices allows for a geometrical efficiency 
of almost 80\% of 4$\pi$,
 also ensuring a  good granularity 
(around 300 electronic channels). 
Both \gar{} and \rc{} are optimized for the detection of  
charged fragments with low energy thresholds. 
The \rc{} is dedicated to 
the detection and identification of forward emitted fragments, which  in the 
presently studied case are mainly   
fragments with Z$\geq$3. They 
are efficiently  identified in charge by the \rc{} via the $\Delta$E-E 
correlation IC-Si and,    
only for fragments with 3$\leq$Z$\leq$8, also using the 
PSA in the Silicon detectors 
with an energy threshold approximatively of 1.5$\,$MeV/\textit{u}.
Light Charged Particles (LPCs) can be isotopically identified 
through Si-CsI(Tl) correlation and PSA in CsI(Tl), while \al{} particles (without mass) 
are identified also using IC-Si correlations and 
PSA in Silicon detectors.
 
On the contrary \gar{} 
ensures the detection of most 
LCPs, which are spread over a wide angular domain. MonteCarlo simulations showed 
that almost the 80\% of the detected LCPs are detected by \gar{}. Since 
for the investigated reaction essentially only LCP fly into \gar{}, in this 
experiment we discarded the \gar{} gas stage and we operated only the CsI(Tl) 
crystals which are good enough for LCP identification and energy 
determination. 
The identification energy thresholds, on average, are 
3, 6, 9 and  5$\,$MeV  for p, d, t and $\alpha$ particles respectively.  
Free neutrons and $\gamma$-rays are not detected. 

The $^{12}$C beam was provided by the XTU TANDEM, at the energy of 
95$\,$MeV with   
an average intensity of about 0.1$\,$pnA: it was pulsed in bunches of 
2$\,$ns width with a repetition period of 400$\,$ns.  

The $^{13}$C target was a self-supporting thin film, with a thickness of 
100$\,\mu$g/cm$^2$.   
The target isotopic purity is known to be more than 99\% at the 
production time.   
During the data taking, probably due to the vacuum level in 
the scattering chamber, the $^{13}$C foil was
polluted by a certain amount of $^{12}$C, which is a common 
contaminant in vacuum systems: indeed,   
residual hydrocarbon molecules are reduced to graphite on the target 
under beam irradiation.   
A similar effect has been described in literature~\cite{bib:papa86, 
  bib:dayras76, bib:kovar79} and recently evidenced in     
another experiment with \gar{}~\cite{bib:csym16}.     
Therefore, similarly to Ref.~\cite{bib:csym16}, it was necessary to 
evaluate this background and subtract its effect from the 
$^{13}$C events.  
 
\section{\label{sec:sel_and_mc}Event selection and \ca{}  background subtraction}  

Experimental results (e.g. Ref~\cite{bib:ortiz}) and model calculations~\cite{bib:gupta, bib:bass} 
show that for this type of system, fusion process accounts for more than 50\% of 
the total reaction cross section.

% Some relevant characteristics of the reaction are summarized on Table 
% ~\ref{tab:reaction}: we observe that more than 50\% of the total reaction cross 
% section is estimated   to produce complete fusion events and this is 
% in agreement with results on similar light systems (e.g. Ref~\cite{bib:ortiz}). 
 
% \begin{table*}[t] 
% \caption{\label{tab:reaction}Some  parameters for the studied 
%   reaction; center of mass  
%   energy, $Q$-value for complete fusion, compound nucleus excitation 
%   energy, grazing angle, maximum fusion angular momentum,  
% total reaction cross section (Gupta model estimation~\cite{bib:gupta}) 
% and fusion cross section  (calculation performed with PACE4 using the 
% Bass model~\cite{bib:bass}).}  
% \begin{ruledtabular} 
% \renewcommand\arraystretch{1.2} 
% \begin{tabular}{cccccccc} 
% \bfseries Reaction & $\mathbf{E_{cm}}$ & $\mathbf{Q_{fus}}$ & $\mathbf{E^{*}}$ & $\mathbf{\theta^{lab}_{gr}}$ & $\mathbf{J_{max}}$ & $\mathbf{\sigma_{R}}$ & $\mathbf{\sigma_{F}}$\\ 
%  & [MeV] & [MeV] & [MeV] & [deg] & [$\hbar$] & [mbarn] & [mbarn] \\  
%  \hline 
%  $^{12}$C+$^{13}$C & 49.39 & 16.31 & 65.7 & 4.9 & 18.5 & 1440 & 780\\ 
% \end{tabular} 
% \end{ruledtabular} 
% \end{table*} 
 
The selection of fusion-evaporation events has been done via software gate,  
requiring the coincidence of only one ``heavy'' ion Z$\geq$5 (the evaporation 
residue, \er{}) with at least one LCP and vetoing the   
possible (rare) coincidences of \er{} with an intermediate mass fragment 
(Z=3,4). 
These latter cases can  be ascribed to the   
break-up channel that is weakly populated   for the excited light 
$^{25}$Mg compound nuclei.  %(E$^{*}$=65,7$\,$MeV),   
Although, these cases are 
interesting and have been recently studied~\cite{bib:manna16} just in 
the context  of possible \al{} cluster effects, we neglected them in this 
paper.  Indeed, we found a branching ratio  for these 
break-up events of less than 1\%, in agreement with the  
PACE4~\cite{bib:pace4_gavron, bib:pace4_site, bib:pace4_lise++}
prediction that  
99\% of the $^{25}$Mg decays are of evaporative kind (with a final \er{} 
Z$\geq$5). %In this work we neglect this minority fission-like decay,

Since the projectile ions have atomic number comparable with that of \er{},
the chosen gate for fusion events can include some background of 
non-central collisions where a quasi-projectile (QP) is detected in coincidence with 
some LCP. To improve the fusion event selection  we 
also required the completeness of the detected charge:   
the sum of the charge of the detected fragments has to be equal to the 
charge of the system, Z$_{sys}$=12.   
The selected sample amounts to only  1.6\%  
(approximatively  3.3~millions) of the total sample, but it represents a high 
quality data set for fusion, with \er{} identified in Z and LCP both in Z 
and A, thus allowing stringent tests on the various decay chains. 
A final requirement has been imposed on the momentum conservation
to remove residual spurious coincidence events, that are anyhow 
overall less than 1~\textperthousand$\,$ within the already selected 
sample; they  are more polluting the channels with a supposed  \er{} in the region of the projectile 
(Z=6$\pm$1). As demonstrated in our previous 
article~\cite{bib:morelli1}, the event sample 
selected in this way does not bias the characteristics of the totality of fusion events.

\subsection{\label{sec:12cpol}Background  of reactions on \ca{}}  

In order to perform a  
detailed analysis of the various evaporation chains, we 
must get rid of the problem of the \ca{} contamination of the enriched 
\cb{} target as mentioned in Sec.~\ref{sec:exp}. 
% First, we demonstrate 
% the presence of this background, then we  
% evaluate the percentage of \ca{} ions in the 
% \cb{} target. 
Reactions on \ca{} ions, indeed, lead to the 
formation of \mga{};  since we are not able to isotopically 
identify the \er{} and we do not measure neutrons,   
the experimental observables for the \mgb{} decays could be 
biased by a spurious contribution of \mga{}. 
Moreover, we are specifically interested in searching weak differences 
between \mgb{} and \mga{} decays, thus   
it is evident the need to keep under control the 
background and to restore a clean  \mgb{} sample of events. 
 
\begin{figure}[b] 
   \centering 
   \includegraphics[width=1\columnwidth]{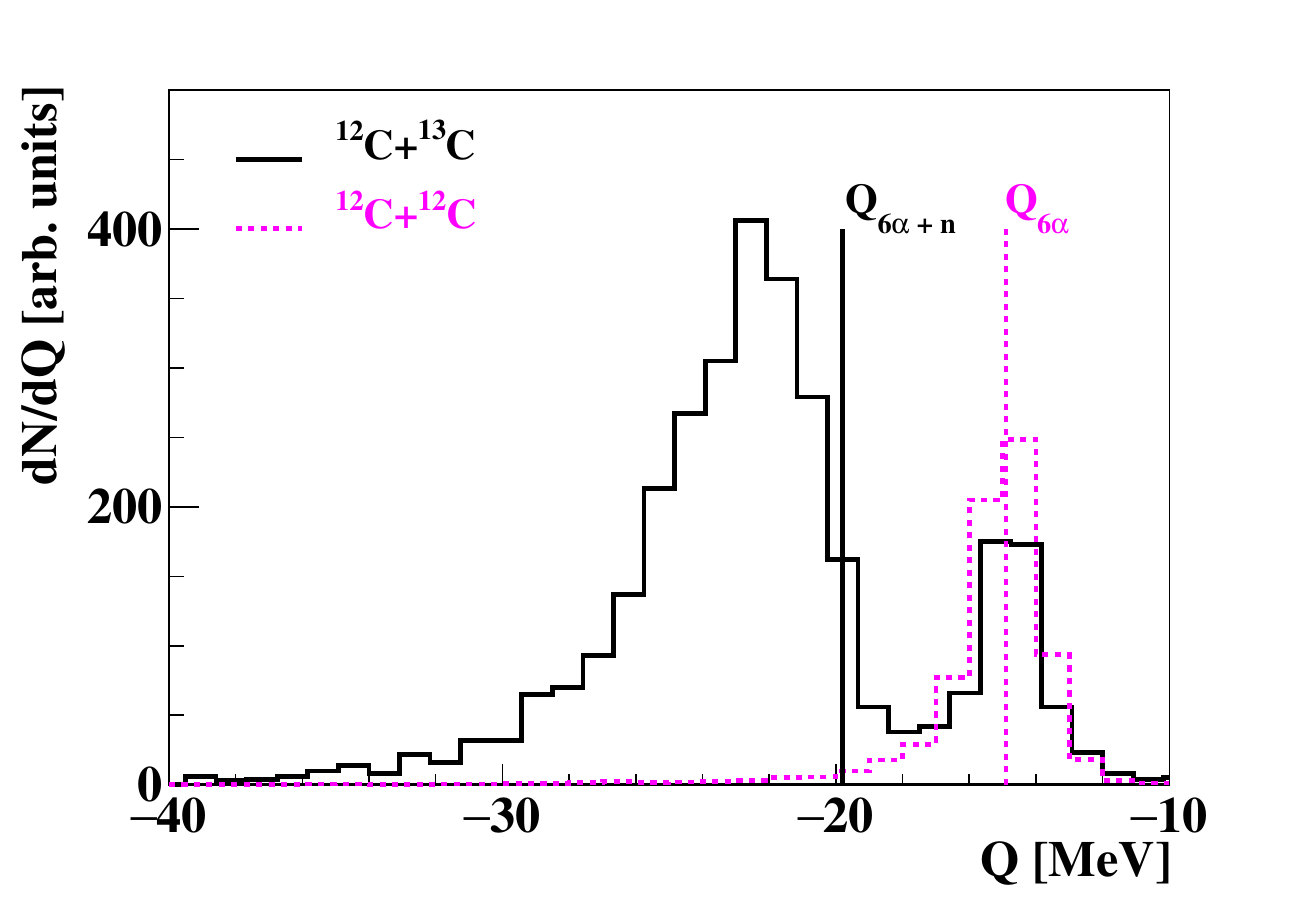} 
   \caption{(Color online) \textit{Q}-value distribution of complete in charge events with six 
     detected \al{} particles. The experimental data referring to the 
     \ca{}+\cb{} reaction are shown as a continuous black line.
     The \textit{Q}-value distribution obtained from the \ca{}+\ca{} reaction is also shown (dashed magenta line). 
     The vertical lines  mark   the expected values 
     for the two channels \mga{} $\rightarrow$ 6\al{} (Q$_{6\alpha}$) 
     and \mgb{} $\rightarrow$ 6\al{}+1n (Q$_{6\alpha +n}$).}  
   \label{fig:hcom_qval6a} 
\end{figure}

In order to evaluate the level of spurious reactions on \ca{} 
nuclei, 
we select a  
specific decay channel where the contribution   
of the background can be easily disentangled, namely the 6\al{} channel. 
We use the reaction 
\textit{Q}-value~\cite{bib:morelli2}:  
\begin{equation} 
\label{eq:qval6a} 
 Q=\sum^{N}_{i}E_{i} - E_{beam} 
\end{equation} 
where $N$ is the number of charged species, $E_{i}$ is the 
lab. kinetic energy of the fragment $i$ and $E_{beam}$ is the beam energy.  
% We evaluated the \textit{Q}-value for the particular  
% channel where the system disintegrates   
% in six \al{} particles. 
Exploiting the fact that \al{} clusters 
have not excited states close to the ground state (the first excited state is approximatively at 20$\,$MeV), 
we should  obtain 
distinct \textit{Q}-values  associated with the disassembly of  
\mga{} in six \al{} particles or \mgb{} in six \al{} plus one neutron  (due to 
fusion with \ca{} or \cb{}, respectively). The spectrum obtained from the runs  \ca{}+\cb{} 
is shown in Fig.~\ref{fig:hcom_qval6a} as continuous line.   

 \begin{figure*}[t] 
   \centering 
   \includegraphics[width=1\columnwidth]{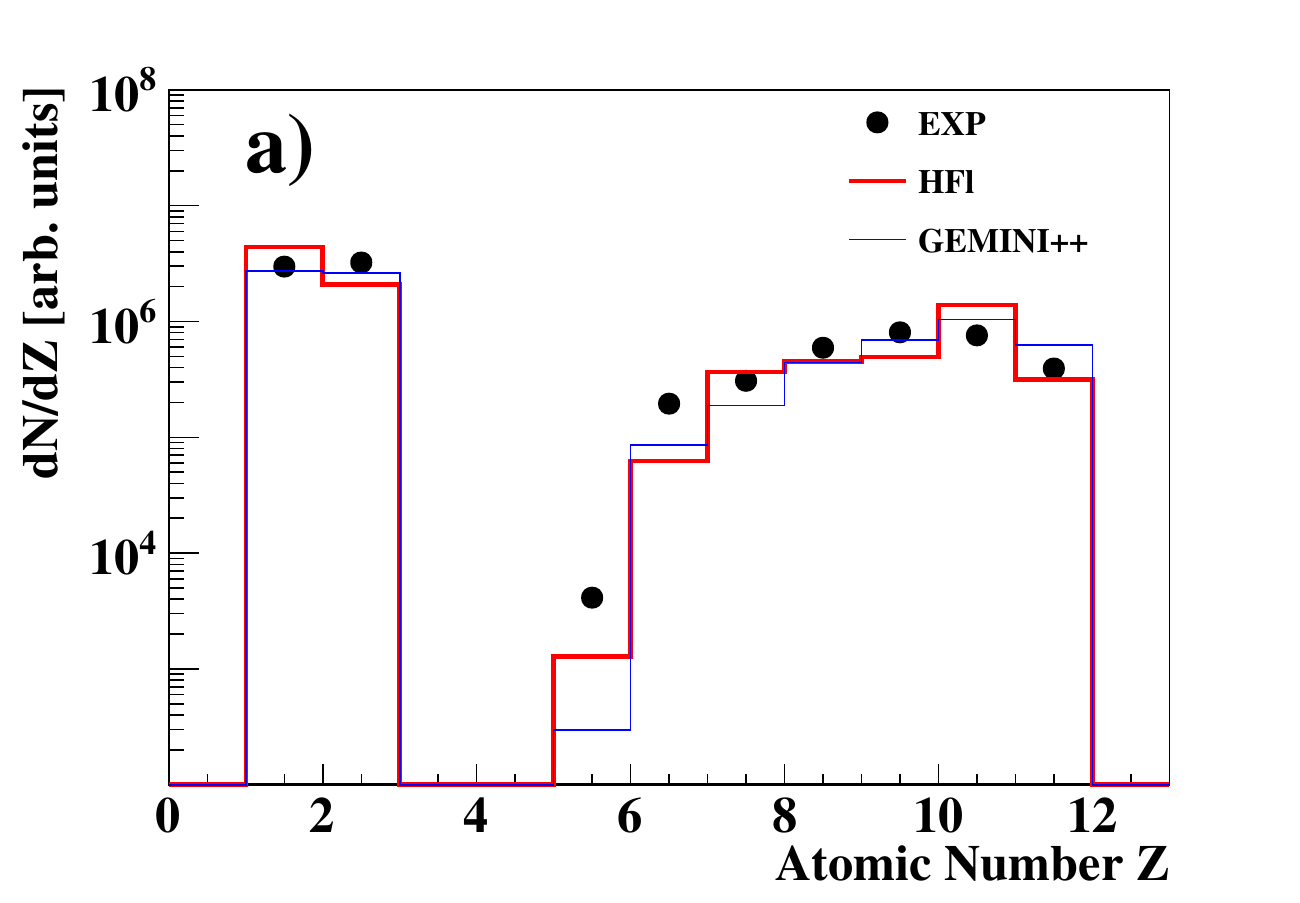} 
\includegraphics[width=1\columnwidth]{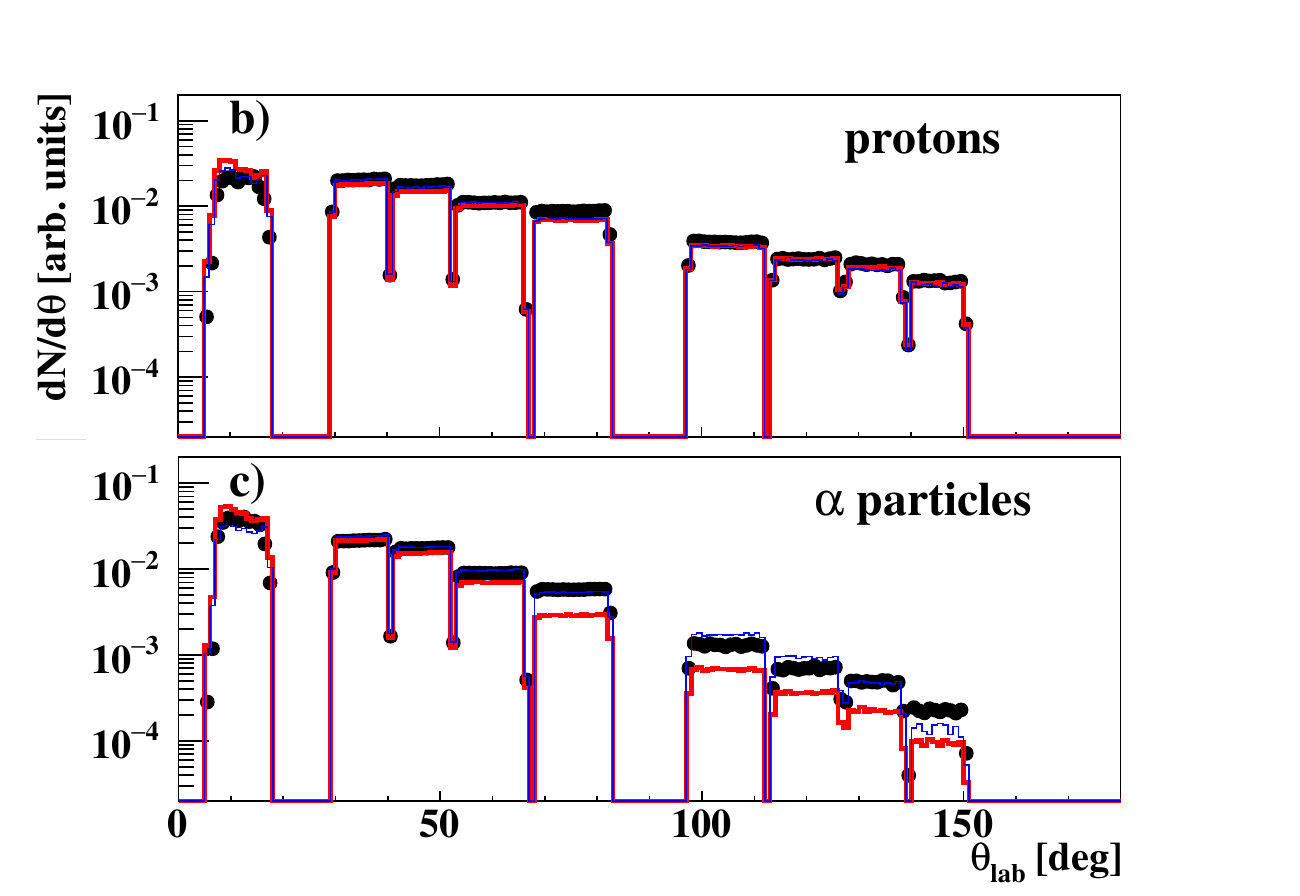}	 
   \caption{(Color online) Part a): charge distribution.
   Part b) and c): proton and alpha angular distributions in the laboratory frame.  
   Black dots are the experimental results and the red bold and blue thin line represent the \hf{} and  
   \gemini{} results, respectively.  Both simulated charge 
   distributions are normalized to the measured number of fusion events 
   complete in charge, while the proton and \al{} angular distributions are normalized to 
   unitary area for a comparison with the results in~\cite{bib:morelli1}.}  
   \label{fig:hz_comp} 
\end{figure*}  

In the figure we observe two distinct peaks.  
The main left peak is rather asymmetric and is upper limited 
by the value Q$_{6\alpha +n}$ with a tail due to the 
missing kinetic energy of the unmeasured neutron. This is the 
contribution of the \mgb{} decay. 
The weaker right peak  is exactly centered at the value expected for  
Q$_{6\alpha}$ associated with \mga{} decay and it
is quite symmetric because no neutrons are missing. 
This result clearly demonstrates the presence of  $^{12}$C+$^{12}$C 
events in the dataset. An 
additional confirmation comes from the  
\textit{Q}~-value distribution obtained for some 
specific $^{12}$C+$^{12}$C runs, 
purposely collected during the same experiment. 
The corresponding peak is  drawn as a dashed (magenta) line in the same 
picture and perfectly matches with
the right peak for  the data on the \cb{} 
target. Moreover, no other peak is present in this case.

A quantitative estimation of the the background due to \ca{}+\ca{}  can be deduced 
as follows.  Directly from Fig.~\ref{fig:hcom_qval6a} we can count the number 
of events related to \mga{} decay into 6\al{} (N$^{6\alpha}_{24\rm{Mg}}$), 
summing the events on the right side with respect to Q$_{6\alpha +n}$ line. 
The percentage of those events is \textit{f}=(18$\pm$1)\% of the total 6\al{} events.
This number is not the real estimation of \ca{} contribution 
because it does not consider the different Branching Ratios (BRs) 
for the \mga{} and the \mgb{} decay into 6\al{}.  
From the analysis of the \ca{}+\ca{} reaction presented in Ref.~\cite{bib:morelli1, bib:morelli2}, 
we estimated that the the \mga{} BR for 6\al{} channel is BR$_{6\alpha}$=3\textperthousand$\,$. 
So the total number of \mga{} events complete in charge, 
N$^{comp}_{24\rm{Mg}}$, can be estimated as N$^{comp}_{24\rm{Mg}}$~=~N$^{6\alpha}_{24\rm{Mg}}$~/~BR$_{6\alpha}$.
Thus the final background level can be obtain dividing this latter by the total number of complete events 
\textit{f'}=N$^{comp}_{24\rm{Mg}}$/N$^{comp}$: 
\textit{f'} results to be (5.8$\pm$0.5)\%. 

With this factor, the  
number of spurious events of reaction on \ca{} can be estimated for (and 
subtracted from) each specific decay chain measured for charge-complete 
events. 
% For instance, the \mga{} experimental distribution in
% Fig.~\ref{fig:hcom_qval6a},    
% obtained from the \ca{}+\ca{} runs and normalized to the same number of decay into 6\al{}, 
% can be  subtracted from the \mgb{} distribution to   
% obtain a clean spectrum.  
Of course,  
it is impossible to disentangle event by event the 
two reactions on $^{12}$C and $^{13}$C. 
Thus \textit{f'} can be used only for an average correction, although 
applicable to all the various decay channels. From now on, we discuss   
spectrum shapes and yields after removal of the $^{12}$C contribution. 
We underline that the $^{12}$C background subtraction
comes out to be very important for the refined analyses presented
in this paper, especially for the reliability of the fit procedure
described in Sec.~\ref{sec:fitmass}.

\section{\label{sec:fus}General characteristics of fusion reactions 
  and comparison with the statistical model} 

Before entering  the detailed analysis of the individual evaporation 
chains, here we present some general features of the fusion event 
class also compared with the model predictions. 

Considering the approach of this paper, it is extremely important to 
perform the analysis of good and selective  experimental data in parallel with 
reliable MonteCarlo statistical codes: as done in 
the past, here we use two different MonteCarlo implementations of the 
statistical model simulations.  
The first one is \gemini{}~\cite{bib:gemini} and the second one is a code 
labelled as \hf{} (\textit{Hauser-Feshbach light}), developed by 
our collaboration~\cite{bib:phd_baiocco, bib:baiocco}  specifically  
for the study of light nuclei  and described elsewhere~\cite{bib:morelli1}. 
Both codes are based on the Hauser-Feshbach 
theory: \gemini{} 
is a general purpose code which includes several  
parameters, continuously refined according to experimental results. In 
particular,  \gemini{} contains  
only a smooth parametrization of the level density, whereas \hf{} 
includes all the known single levels as found in the available 
databases~\cite{bib:nudat2}.   
This feature is important for the detailed investigation of the 
various decay channels, as we want to pursue.  
% An other difference between the two models concerns the spin 
% treatment. \gemini{}, being a general use code, accepts only integer 
% values for the compound nucleus spin; \hf{}, instead, including 
% details on the nuclear structure, allows also for semi-integer spin 
% values and this option of course makes the use of \hf{} more proper for 
% this work (the g.s. of \mgb{} has a spin 5/2$\,\hbar$). 
 
For both simulations we choose, as  default option, a triangular distribution 
for the angular momentum of the compound nuclei extending from zero to 
the ma\-xi\-mum angular momentum for fusion  
($J_{max}$=18.5$\,\hbar$ calculated with  
the code PACE4 using the Bass model~\cite{bib:bass}).  
For the spin distribution  we 
fixed a tail with    
a diffuseness parameter $\Delta J$=2$\,\hbar$ as proposed for  
similar light systems~\cite{bib:dey2009, bib:mahboub} and with peaking 
values J=15.5$\,\hbar$. 
The shaping recipe for the spin distribution 
is quite similar to that 
of our previous paper~\cite{bib:morelli1} for the $^{12}$C+$^{12}$C system.  
Since the  spin distribution of the studied CN is not exactly known, 
we verified the model predictions as a function of reasonable changes 
in the spin values. This will be discussed afterwards but we anticipate that 
the conclusions are essentially unaffected by different assumptions on 
the spin.   
  
The simulated events have then been filtered via a software replica of the apparatus,  
including the efficiencies, the resolutions  and the 
identification thresholds of the various detectors.   
Of course, the same  selection criteria to identify fusion events imposed on 
measured data have been applied as well.

In Fig.~\ref{fig:hz_comp} (a) the charge distribution for complete events  is 
shown for the experimental and simulated events selected through the fusion 
gates.  In the following, unless otherwise noted, \hf{} and \gemini{} 
results are reported in red bold and blue thin line respectively.  
We clearly observe, as expected, the bell-shaped region of the \er{} 
(Z$\geq$5) well separated from the LCP part. We remind that the absence 
of Z=3,4 ions is only due to the selection gate  chosen for 
this analysis.  
 
The overall behaviour of the experimental distribution  
is well reproduced by both simulation codes.
% :  
% this somehow confirms that we are indeed properly selecting  
% the fusion-evaporation channel of $^{25}$Mg compound nuclei.  
However, some discrepancies between experimental and simulated results 
appear in particular for the relative abundances of LCP. Both models 
underestimate the emission of Z=2; while only \hf{} overestimates the emission of Z=1. 
Looking at the \er{} region of the fragment distribution, instead, we can 
observe that statistical models (more \hf{} than \gemini{}) nicely 
reproduce the yield of odd Z residues, while some disagreement is found 
for even charge residues (mostly Z=6,8). 
In the following it will be shown how these discrepancies, observed in 
detail, are related to the 
emission of only $\alpha$ particles from the compound nucleus,   
since even-Z \er{} can be reached  through the evaporation of only 
$\alpha$ particles while  odd-Z \er{} chains need the emission of at least one 
hydrogen ion.  
 
Complementary information can be seen in the other panels of Fig.~\ref{fig:hz_comp}, 
where the angular distributions for proton and \al{} particles  are shown in the laboratory frame (part b) and c) respectively).  
The angular range below 20$^\circ$ is covered by the \rc{}, while particles 
above 30$^\circ$ are detected in the \gar{} CsI crystals.
% Within each detector, 
% a uniform random spread of the impinging particles on the active surface of the detector 
% has been applied.
The shown angular distributions are normalized to unitary area  
for shape comparison. Both \gemini{} and \hf{} follow the  
experimental proton distribution at all emission angles. 
Moreover, also the angular distributions of deuterons and tritons (which 
altogether represent a minority fraction of 7.5\% of LCP)  are  
well reproduced by both models; they are not shown for brevity. 
The \al{} angular distribution, instead, shows a favored  emission 
at more backward angles with respect to \hf{}, while is in  quite good agreement 
with the \gemini{} prediction. 
Comparing our data with those obtained for the \mga{} decay, the same
behaviour   
is observed as shown in Fig. 7 of~\cite{bib:morelli1}. 

As previously mentioned, the exact shape of the spin distribution of 
the CN is not known. On the other hand, within statistical models, the 
particle production rates and their phase-space properties are 
somewhat dependent on the assumed spin values~\cite{bib:dey2009, bib:mahboub}. 
Therefore, we explored how the statistical model  
predictions 
are affected by variation of the spin 
parameters. We run two additional calculations, assuming  for both  
a sharp cut-off spin distribution. For the first run  we put 
the cut-off at  $J_{max}$=15$\,\hbar$, being aware that it would  
underestimate the fusion cross section; for the second set we kept the  
limit value of the default simulation  
 $J_{max}$=18$\,\hbar$ (and $\Delta J$=0$\,\hbar$)

Some differences appear for \gemini{}. In 
particular, the use of the lower value $J_{max}$=15$\,\hbar$ decreases the 
\al{} emission (-7\%) while it slightly increases the proton yield (+7\%), thus further 
enlarging the differences with the experimental data.  
For \hf{}, instead, the effects are less than  1\%. 
The effect of 
zeroing  the diffuseness is almost negligible, both for \gemini{} and \hf{},  
increasing by only  1\% the \al{} particles yield. 
Having proved that reasonable modifications  on the spin distribution 
do not reduce significantly  the observed discrepancies at this level of analysis, we 
 choose the default calculations (defined at the beginning of this section) as reference and using the other simulations to estimate the 
 systematic uncertainties on the final results (see Sec.~\ref{sec:dev}). 
 
\begin{figure*}[t] 
   \centering 
   \includegraphics[scale=0.6]{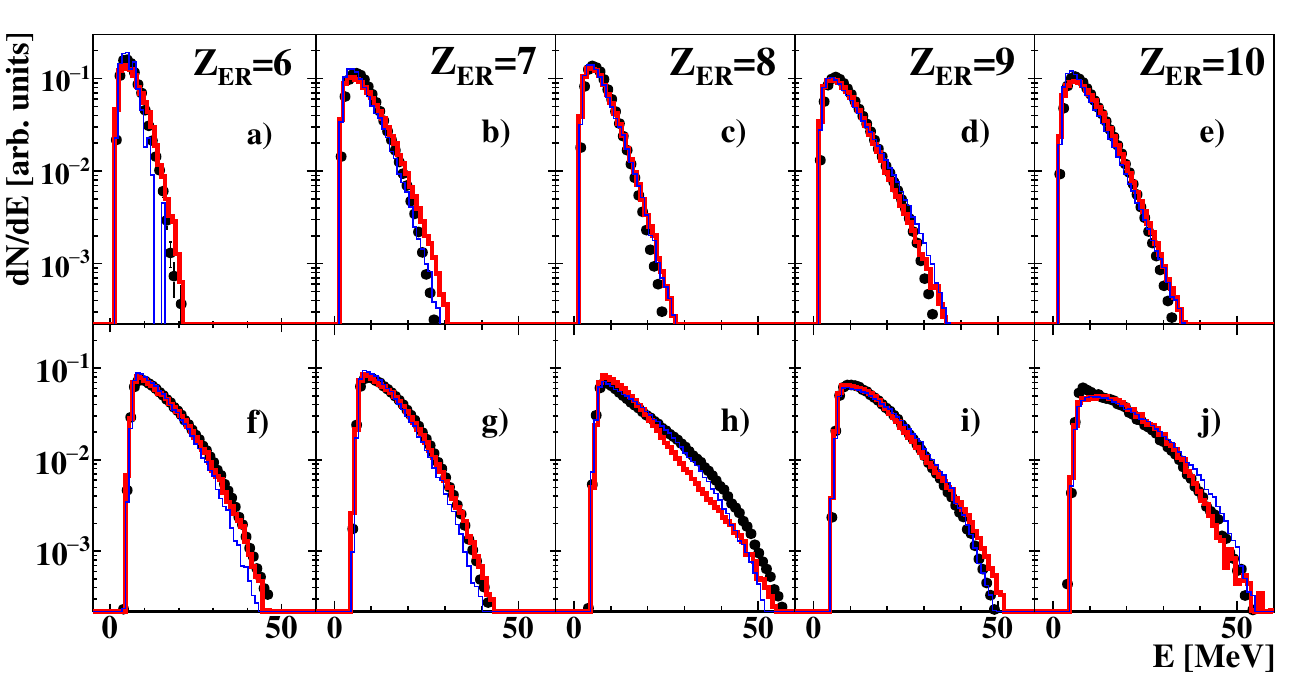} 
   \caption{(Color online) Kinetic energy distributions in the la\-bo\-ra\-tory frame for 
     protons (from a) to e) panels) and $\alpha$ particles (from f) to l) panels) 
     identified in \gar{}. Points represent experimental data, thin blue
     and bold red histograms the \gemini{} the \hf{} predictions respectively.  
   The distributions refer to different coincident \er{}, 
   from Carbon to Neon and are normalized to unitary area.} 
   \label{fig:hcom_paER} 
\end{figure*}
 
\section{\label{sec:dev}Results for selected decay channels} 

In Fig.~\ref{fig:hcom_paER} the kinetic energy distributions in the 
laboratory frame  
for protons (from a) to e) panels) and $\alpha$ particles (from f) to l) panels) 
detected  with \gar{} in 
coincidence with \er{} from Z=6 up to Z=10 are shown.  The collected
statistics of the channel with Z$_{ER}$=5 is not enough   for this
analysis.  The experimental  
results are indicated with filled dots while the lines are the model 
predictions (see caption for details). The distributions are  
normalized to unitary area for a easier shape comparison. 

At a first sight we see that the models nicely follow the 
experimental data for all channels. Going into details of the various chains, 
for protons we observe a very good agreement between the experimental and 
si\-mu\-lated shapes, with both models. Instead, some 
differences appear for \al{} 
particles especially for the chain ending with Z$_{ER}$=8 (Fig. \ref{fig:hcom_paER} h))
where the  measured high energy tail is  not well reproduced by the 
models. There is also some disagreement in the \al{} spectra associated with 
Z$_{ER}$=6,7 (f) and g) respectively) with respect to 
\gemini{}, while \hf{} better follows the data.  The simulated shapes are 
negligibly affected by (reasonable) changes of the parameters 
ruling the CN spin distribution.  
The good success of the statistical models makes us confident about the 
investigation of the further details. 
The discrepancies of the measured LCP multiplicities with 
respect to model  
predictions (see Sec.~\ref{sec:fus}) together with the differences in the 
\al{} energy spectra for specific evaporation chains, are similar to 
the findings of 
Ref.~\cite{bib:morelli1} on \mga{}; in this latter case a slight shape difference between the measured and \hf{} simulated energy 
spectra of \al{} particles was found also for Z$_{ER}$=6, not visible in the present data (Figure~\ref{fig:hcom_paER}). 
In order to explain such differences,  
it was argued there that \al{} emission could be favored, with respect to 
statistical models, for those channels where only \al{} particles are 
emitted. In turn, this could be an indication for some non-statistical effects,
not included in  the Hauser-Feshbach formalism.

\subsection{\label{sec:oxy}The case of the Oxygen evaporation 
  residues}

We now focus on the 
Oxygen channel presenting the biggest anomalies and then on the 
branching ratios of the channels dominated by \al{} particle emission. 

\begin{figure}[h] 
   \centering 
   \includegraphics[width=1\columnwidth]{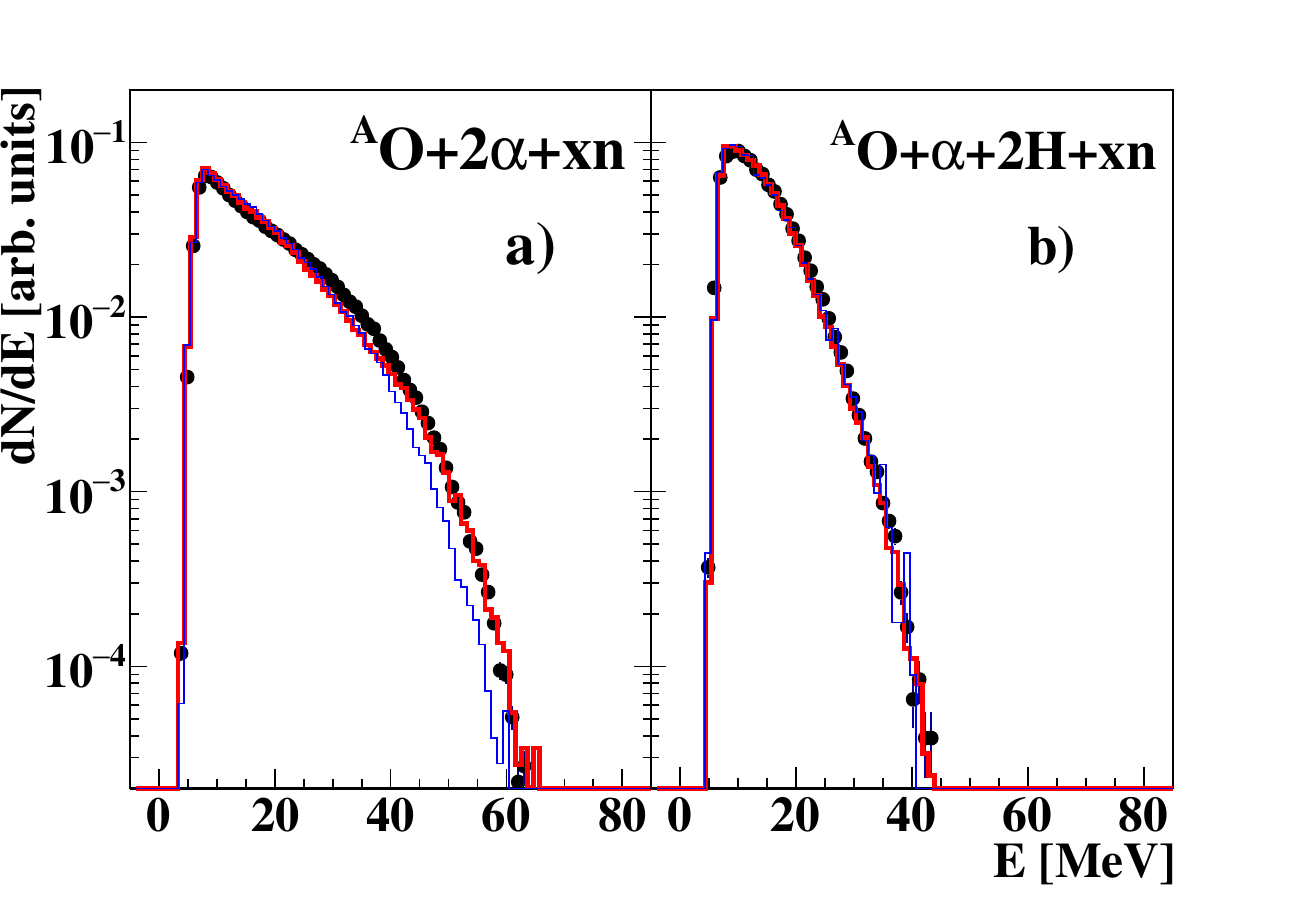}\\ 
   \includegraphics[width=1\columnwidth]{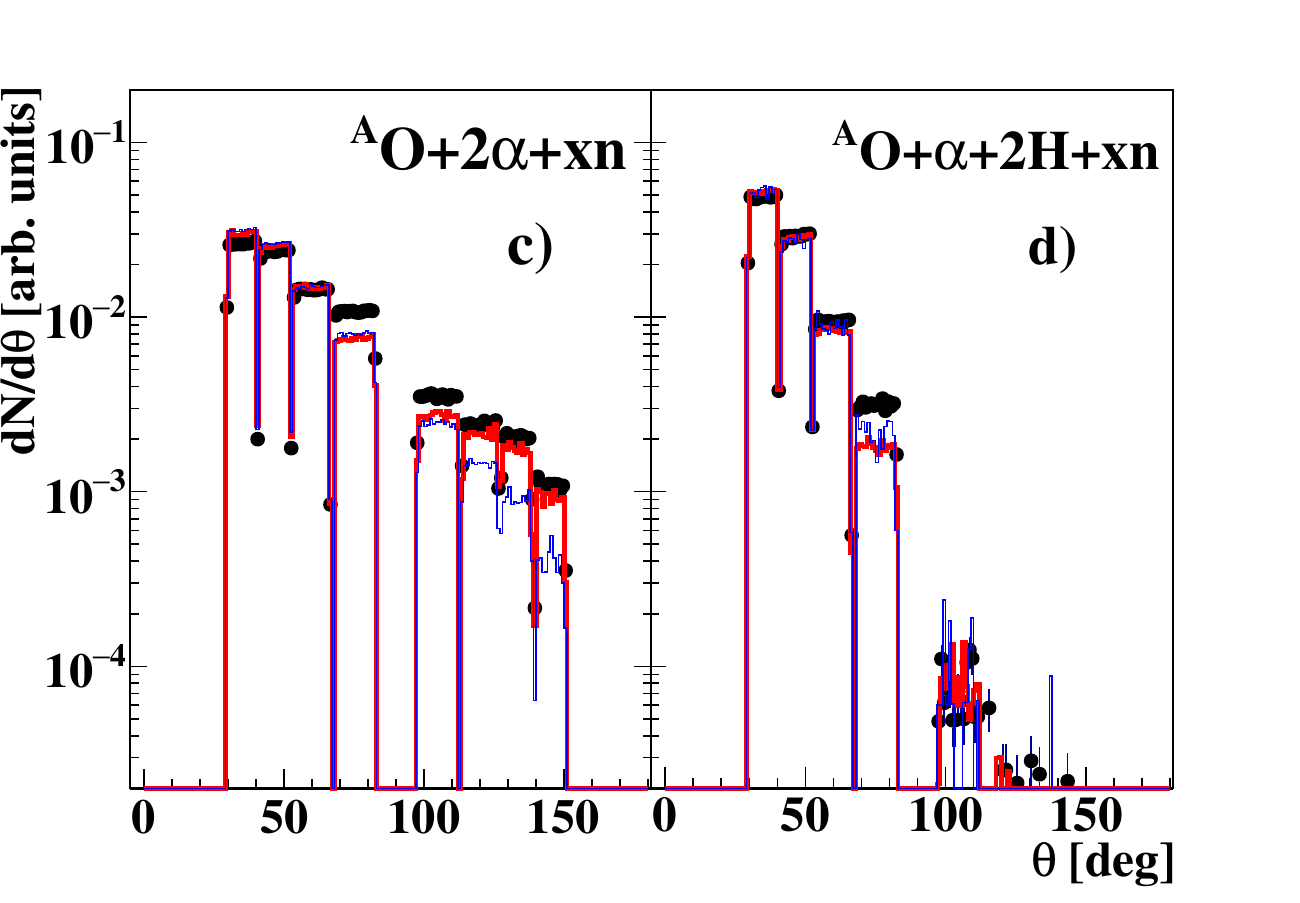} 
   \caption{(Color online) Lab. kinetic energy (a) and b)) and 
     angular (c) and d)) distributions for $\alpha$ particles for 
     the channel with an Oxygen \er{}. Experimental and 
     si\-mu\-la\-ted results are drawn according to the convention of the previous figures. 
   The two main contributing decay channels are considered, in particular 
   $^{A}$O+2$\alpha$ (a) and c)) and $^{A}$O+$\alpha$+2H (b) and d)).  
   The distributions are normalized to unitary area.}  
   \label{fig:hcom_oxy} 
\end{figure}

In Fig.~\ref{fig:hcom_oxy},  
the \al{} particle energy and angular distributions for the two chains 
 $^{A}$O+2$\alpha$ (panels a) and c)) and $^{A}$O+$\alpha$+2H (where H 
means Z=1) are shown, each normalized to unitary area. 
These are the two chains mainly  contributing to 
Oxygen production. The experimental and simulated data are drawn 
according to the already introduced convention. 
The results for the $^{A}$O+$\alpha$+2H 
chain (right part) are fully compatible   
with the statistical model predictions as shape. 
A quite good agreement is found also for 
the channel $^{A}$O+2$\alpha$, but   
only using the \hf{} code; \gemini{} less faithfully follows the 
experimental energy and 
angular distributions. 
 
The high level of accuracy of the \hf{} calculations in reproducing the 
phase-space of emitted LCP (over more than three orders of magnitude) 
demonstrates the importance 
of including in the model as many details as possible of 
the nuclear structure for the relevant nuclei. Due to 
this improvement of \hf{} with respect to \gemini{}, in the  rest of the paper 
we will limit the comparisons to the \hf{} code only. 
 
From Fig.~\ref{fig:hcom_oxy}, we can state that 
the kinematics of the chains ending up with an   
Oxygen \er{} is accurately  reproduced by a pure statistical model (\hf{} code). 
However, since global LCP multiplicities and some inclusive $\alpha$ 
particle energy distributions (Fig.~\ref{fig:hcom_paER})   
show deviations with respect to the predictions, one can deduce that   
the weights of the various chains are not fully accounted for by 
the model. In other words, we must verify the quality of the  
model predictions as far as the BRs for the various 
channels are concerned, just as done in Ref.~\cite{bib:morelli1}.  

\subsection{\label{sec:brs}The branching ratios of the various chains} 
 
In Table~\ref{tab:brs} we report the BRs for the most probable chains containing the largest allowed \al{} multiplicities, for each  Z$_{ER}$.
The contributions due to the different \er{} isotopes which correspond to a different number of emitted 
neutron (x), are summed. Each BR is normalized to the total number of complete 
events with the same Z$_{ER}$.
The errors of the experimental BRs reflect the uncertainties due to 
spurious $^{3}$He in the \al{} identification gates. 
The experimental BRs are compared with the \hf{} results. For the model 
we quote  ranges as fiducial limits of the BR when 
changing the CN spin distributions as explained 
in Sec.~\ref{sec:fus}. 
Statistical errors are negligible. 
 
\begin{table}[t] 
\caption{\label{tab:brs}Branching ratios for relevant evaporation 
  chains. Experimental and \hf{} predictions are compared. Only the 
  most probable chains  with the largest possible \al{} multiplicities, for a fized \er{}, are considered. 
  Errors on the   
  experimental values take into account the possible $^{3}$He-$\alpha$ 
  contamination, estimated to be around 4\%.   
  The model ranges are to consider the effect of the poor 
   knowledge of the CN spin distribution. Statistical 
  errors are negligible in all cases. 
  All the values are normalized to the number of event for each Z$_{ER}$.} 
\begin{ruledtabular} 
\renewcommand\arraystretch{1.2} 
\begin{tabular}{cccc} 
\bf Z$_{\mathbf{ER}}$ & \bf Channel &\bf EXP [\%] &\bf \hf{} [\%]\\%&\bf \gemini{} [\%]\\ 
\hline 
 %11 & $^{24-x}$Na+$x$n+p & 83$\pm$2 & 92 & 82 \\ 
 10 & $^{21-x}$Ne+$x$n+\al{} & 29$\pm$1 & 3.2$\div$3.8 \\%& 30$^{+5}_{-11}$ \\ 
 9  & $^{20-x}$F+$x$n+p+$\alpha$ & 86$\pm$3 & 84$\div$86 \\%& 91$^{+1}_{-1}$ \\ 
 8  & $^{17-x}$O+$x$n+2$\alpha$ & 69$\pm$3 & 30$\div$32 \\%& 80$^{+2}_{-10}$ \\ 
 7  & $^{15-x}$N+$x$n+p+2$\alpha$ & 83$\pm$3 & 90$\div$92 \\%& 84$^{+1}_{-1}$  \\ 
 6  & $^{13-x}$C+$x$n+3$\alpha$ & 97$\pm$4 & 79$\div$83 \\%& 99$^{+1}_{-1}$ \\   
\end{tabular} 
\end{ruledtabular} 
\end{table} 
The most important observation is that the model quite nicely reproduces  
the BRs of the chains containing  an evaporated hydrogen isotope but 
it misses the BRs for pure \al{} emission channels (plus possible neutrons). 
In particular, we find that for these channels \hf{} con\-si\-de\-ra\-bly
underestimates the BRs with relative difference which decreases
 increasing \al{} multiplicity, in agreement with what observed in Ref.~\cite{bib:morelli1}
(see Table$\,$1 in that paper); here the effect is  smaller in
magnitude, except for the Ne+\al{} channel.  
This failure, in the case of  
Oxygen residues,  explains the differences in the \al{} energy spectrum
seen in  Fig.~\ref{fig:hcom_paER} which look like the ones 
for the $^{24}$Mg
(in particular, Fig.13 of Ref.~\cite{bib:morelli1}). 
 
Therefore, the additional neutron of \mgb{} with respect to $^{24}$Mg seems not to strongly modify the decay paths, at least in 
this fusion reaction where the CN has a relatively high excitation energy:    
also for the $^{25}$Mg, indeed, the channels involving the evaporation 
of only $\alpha$ particles result to be favored with  respect to what 
predicted by a pure  statistical model. 
 
\section{\label{sec:confr}More refined comparison between the decays 
   of $^{25}$M\lowercase{g}  and $^{24}$M\lowercase{g}} 
  
In order to further investigate   the   
%these more abundant  
\al{} evaporation chains from  
%in the decay of  
excited Mg nuclei,   
we can directly compare the results obtained for the two fusion 
reactions forming \mga{} and \mgb{}. 
This comparison is quite effective because the 
data have been collected with the same apparatus and with similar 
analysis criteria; therefore possible systematic errors should 
poorly affect  this comparison. 

It would be very  interesting to select 
the evaporation paths on the basis of the emitted neutron 
and, possibly, its emission order,  for the \mgb{}. 
In this respect, valuable  information can be gained by the analysis 
of the \textit{Q}-value distributions. Indeed these distributions, for 
Z-constrained events, contain some footprints of the  
evaporated neutrons. 
\begin{figure} [t]
   \centering 
   \includegraphics[width=1\columnwidth]{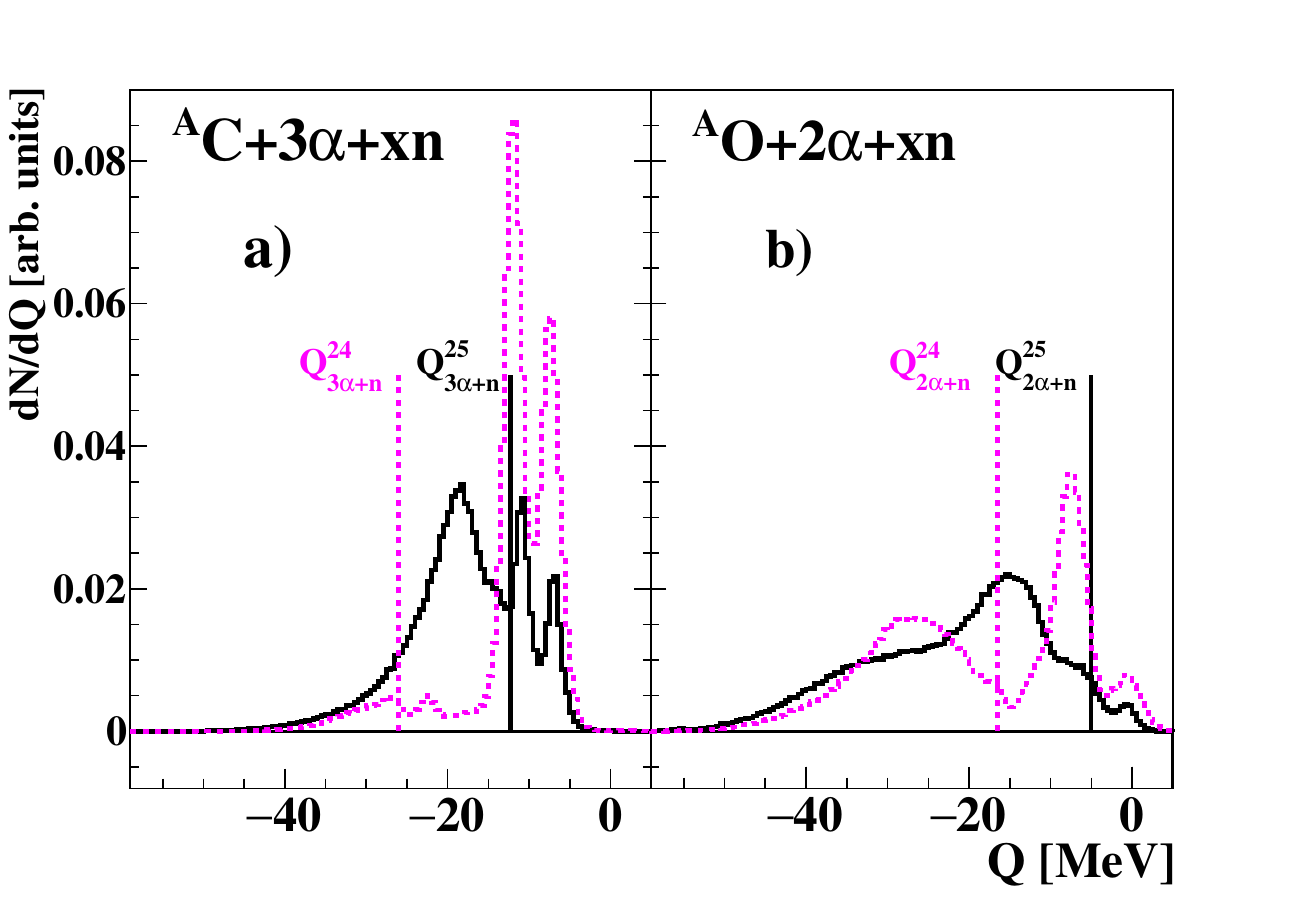} 
   \caption{(Color online) \textit{Q} value distributions for the decay 
     channel C+3$\alpha$+$x$n (part a)) and O+2$\alpha$+$x$n (part b)), for the $^{25}$Mg and $^{24}$Mg  drawn 
   with black continue and magenta dashed line, respectively.  
   The vertical lines are in correspondence of the one neutron
   emission      threshold for each system.  
   The two distributions are shifted by the amount of the neutron separation energy $S_{n}$ in \mgb{}.
   }   
   \label{fig:hqval32a} 
\end{figure} 
In Fig.~\ref{fig:hqval32a} we present  
the experimental \textit{Q}-value distributions  (see eq.~\ref{eq:qval6a}) 
for the two example  chains $^{A}$C+$x$n+3$\alpha$ (part a)) and 
$^{A}$O+$x$n+2$\alpha$ (part b)), for the two compound nuclei \mgb{}
(continuous black line) and 
$^{24}$Mg (dashed magenta line). All curves are normalized to unity. 
Moreover, for a better comparison, the \mga{} distribution has been
shifted in order  to match the \mga{} reaction \textit{Q}-value with
that of \mgb{} case (so  that the ground-state values are aligned). 
In the pictures the  vertical (continuous and dashed) lines correspond 
to the (one) neutron emission threshold for each system. Therefore,
events on the right-hand side of the 
marks are neutron less and end up at the heaviest possible \er{}, either in its 
ground or excited (but particle bound) states. In these latter cases 
the \textit{Q}-value peaks at the energies corresponding to the 
emitted (and undetected) $\gamma$-rays. Instead, in events on the 
left-hand side of the marks at least one   
neutron has been emitted. Since neutrons are undetected, the description of  the low-\textit{Q} 
region of the distributions is not easy because the energy balance is 
incomplete; an original and more accurate analysis of these  
distributions will be discussed in Sec.~\ref{sec:fitmass}.  

For the 3\al{} decays, the rightmost peaks around $-7.3\,$MeV correspond to the ground 
state of \ca{} and \cb{}, respectively for the \mga{} and \mgb{} cases. 
The second peaks from the right are due to the population of the first 
Carbon excited states; there is a single line at 4.4$\,$MeV   
($^{12}$C) for \mga{}, while,    
for the \mgb{} case, we observe a mixed structure due to the
three lower levels of  $^{13}$C  (3.0, 3.6, 
3.9$\,$MeV), not energetically resolved. The small peak 
around $-23\,$MeV visible in the \mga{} case  
is due to a spurious contribution   
from the channel $^{13}$C+$^{3}$He+2$\alpha$~\cite{bib:morelli2}. 

For the Oxygen-2\al{} channel (right-hand side of Fig.~\ref{fig:hqval32a}), 
the events ending with an Oxygen in the ground state are located at $-0.9\,$MeV;  
for the \mga{} distribution the peak around $-7\,$MeV corresponds to
events  where $^{16}$O
is populated in the first excited state (6$\,$MeV). In the case of 
\mgb{} no clear structures associated with excited states of $^{17}$O can 
be seen, also due to the finite energy resolution.
As a general comment, we can note that  for  these
  channels, ending at Carbon or  
Oxygen residues through the emission of  \al{} 
particles, the probability to have 
additional emitted neutrons is larger for the \mgb{} than for the \mga{}
case. Indeed, the  
relative yield beyond the neutron emission threshold is evidently  larger for 
\mgb{}. This means that in these cases, after the neutron emission the
two decay paths resemble each other and, thus, reach the same \er{}.

In order to further separate the various decay chains and obtain a more
stringent comparison, we now try
to reconstruct also the \er{} mass by exploiting  
the shape of the \textit{Q}-value distribution.

\subsection{\label{sec:fitmass}Mass reconstruction in selected decay chains} 

%   We already observed (Sec.~\ref{sec:confr}) that  
%   the \textit{Q}-value (eq.~\ref{eq:qval6a}) distributions 
%   contain information on neutrons, because their 
%   shape is affected by the presence of one or more neutrons emitted 
%   along the chain.
%   In the following we present a method to reconstruct the \er{} mass starting from the \textit{Q}-value distribution.

  \begin{figure}[t] 
   \centering 
   \includegraphics[width=1\columnwidth]{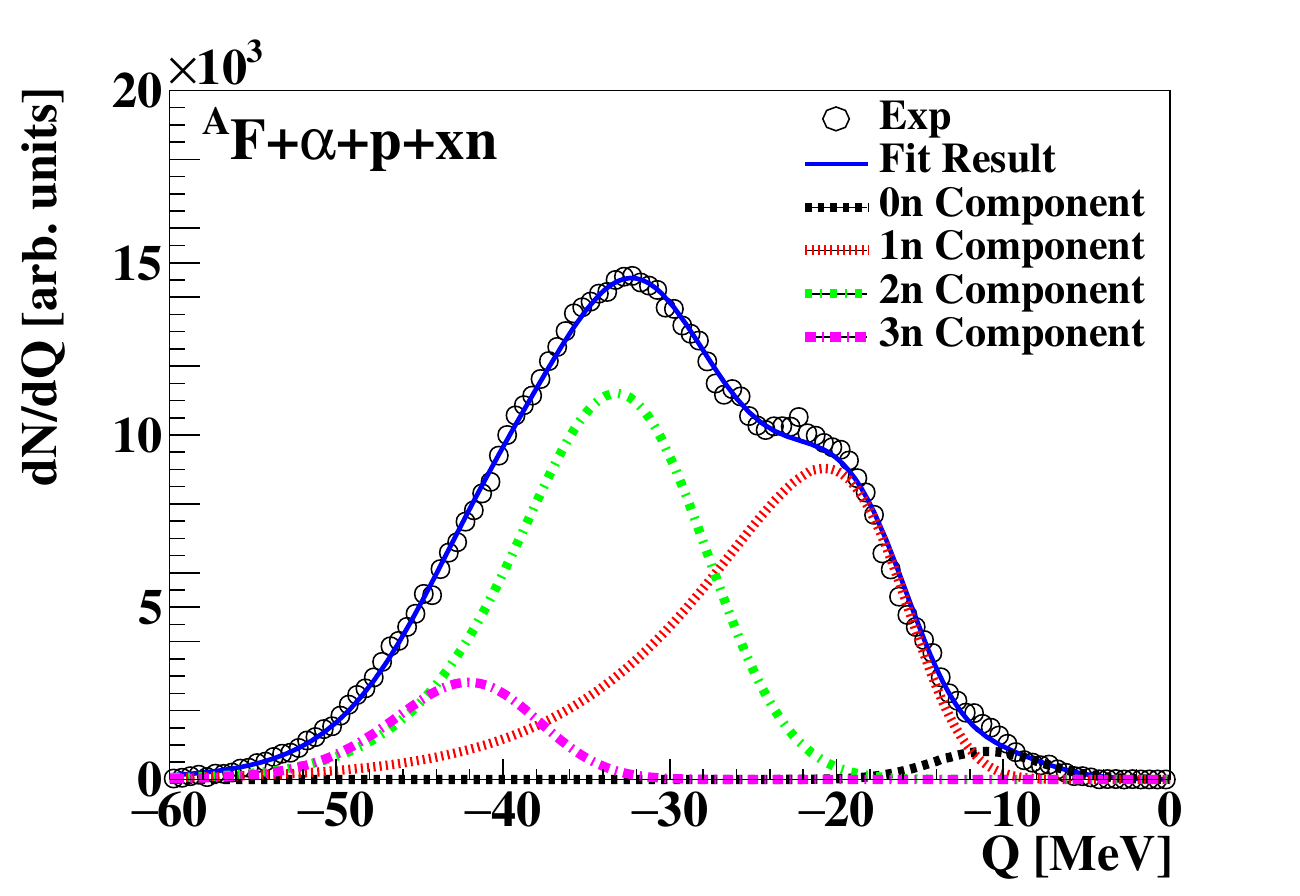} 
   \caption{(Color online) \textit{Q}-value experimental distribution for
     the $^{25}$Mg$\rightarrow ^{A}$F+p+$\alpha$+$x$n chain (open dots).
     On the figure also the result of the fit procedure is shown 
     (blue line) which     is the sum of the various contributions 
     related to the different neutron multiplicities represented as
     explained in the legend.} 
   \label{fig:fit} 
\end{figure}

  As previously discussed, only  below the neutron  emission threshold we
  know exactly the \er{} mass because the neutron multiplicity ($m_{n}$) is zero.
  Above this threshold the 
  shape is modeled by the kinetic energy taken by the emitted neutron; in 
  some cases (i.e. for some specific evaporation paths) more than one 
  neutron can be present and therefore \textit{Q}  extends to even lower 
  negative values due to the larger energy deficit. Of course, for each 
  emitted neutron, the mass of the final \er{} isotope is reduced by one unit. 
  For each selected evaporation path, defined by  a given \er{} 
  and its accompanying LCPs, we attempted to reconstruct the isotopic 
  population of the \er{} through a convolution fit of the 
  \textit{Q}-value distribution.
  
  For the fit we need to fix some 
  functional forms and parameters. The functional forms of the n-fold 
  neutron emissions have been modeled on the basis of the statistical 
  model. Indeed, here we can select chain by chain the various \er{} 
  isotopes and study the shape of the \textit{Q} distributions as a 
  function of the neutron multiplicity. Basically, we adopted 
  two different functionals for $m_n=0$ and $m_n>0$. In the former case, 
  we assumed a Breit-Wigner function convoluted with a Gaussian to keep into account the energy 
  resolution: the initial  widths are suggested by the 
  \hf{} simulation (mostly affected by experimental resolution) and centered at 
  the known energy levels of the \er{}. In the case 
  of neutron emission, each n-fold neutron contribution has the shape of the 
  convolution of 
  a Gaussian and a Maxwellian,  whose defining parameters are tuned 
  basing on the MonteCarlo results. The Maxwellian distributions start from the
  n-fold emission thresholds towards lower values of \textit{Q}. 
  The fit is applied to the measured distributions, for each type of 
  chain constrained in \er{} charge and LCP. 
  The relevant free parameters are the 
  weights of the various n-fold neutron contributions from which we 
  can then reconstruct the \er{} isotopic distributions. 
  
  Using this method we can reanalyze the chains of  Table~\ref{tab:brs}.
  An example of the high quality of our fit procedure is shown in
  Fig.~\ref{fig:fit} for the chain
  $^{25}$Mg$\rightarrow ^{A}$F+p+$\alpha$+$x$n path. The experimental distribution is
  shown (open dots) together with the  fit result (blue line) which
  is the sum of the various components related to
  different neutron multiplicities,  also shown in the picture.
  Errors have been computed varying the slope of each  Maxwellian 
  distribution by 10\% around the estimated value.

\begin{figure}[t] 
   \centering 
   \includegraphics[width=1\columnwidth]{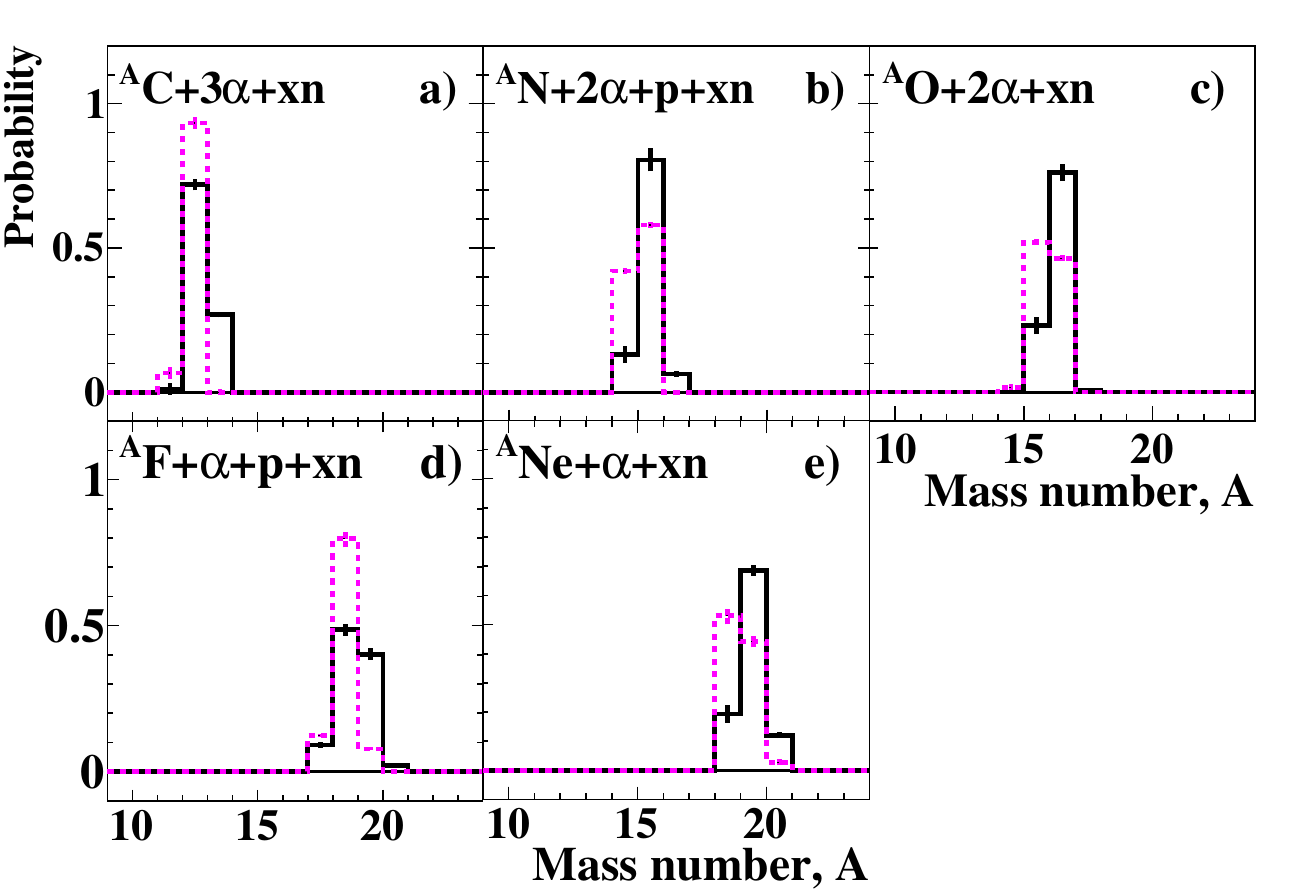} 
   \caption{(Color online) Mass distributions for various \er{} reached
     in the chains Tab.~\ref{tab:brs}. The \mgb{} (black continuous line) case is compared to the 
     \mga{} case (magenta dotted line). All histograms are normalized
     to unitary area.}
   \label{fig:isotopi_soloa} 
\end{figure}

    The weights obtained from the fit allow to deduce the \er{} mass
    distributions. The results for \mgb{} and \mga{} are compared in 
Fig.~\ref{fig:isotopi_soloa}, drawn with continuous and dotted lines,
respectively. 

We note that the initial larger $N/Z$ value of
  the source in the case of \mgb{} brings to slightly heavier
  \er{}. Indeed, the average mass    for each \er{} charge value is
  0.3-0.4~amu larger. However the shift is lower than one amu  implying that 
  in most cases  the additional neutron is not emitted 
  as the first particle in the decay chain.
  We note that \hf{} simulations predict average \er{} masses which agree with the
  measured ones within 20\%.
  In particular, if the extra neutron is 
  preferentially emitted in the first evaporation step, 
  the detected events would correspond to the decay of a \mga{} source, 
  which would explain why the results are similar to the ones of Ref. \cite{bib:morelli1}.  
  However, the results of Fig.~\ref{fig:isotopi_soloa} 
  do not allow to discriminate between the different emission steps.
Thus we explore some other variables  describing phase-space  
  correlations  among \er{} and emitted  particles and possibly 
  sensitive to their emission order.

\section{\label{sec:jacobi} Emission pattern for the Oxygen-2\al{} channel}

 \begin{figure}[ht]
   \centering
\includegraphics[width=1\columnwidth]{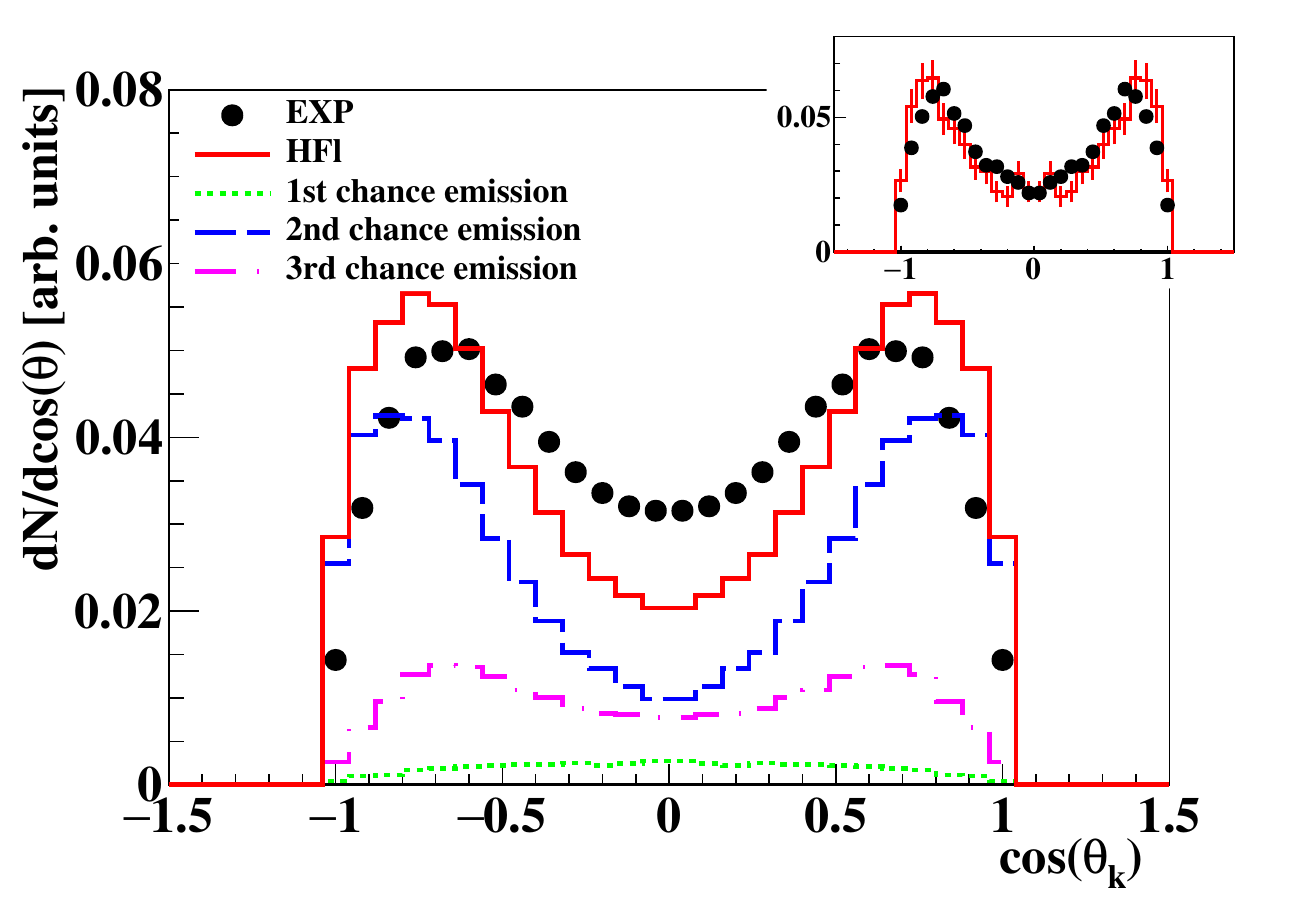}
   \caption{(Color online)
   Probability distribution of $\cos(\theta_{k})$ for experimental (black dots)
   and \hf{} events of 
   the type \al{}-\al{}-1n-Oxygen: both distribution are normalized to unitary area. 
   The figure also shows the  various
   cases corresponding to the three emission orders
   of the neutron (see legend).
   The sum of these three cases gives the total \hf{}
   curve (red continuous line). The weights 
   are those predicted by \hf{}.
   Each contribution is scaled by its
   weight to show the relative contribution to the total \hf{}
   distribution, before     and after the fit procedure.
   In the sub pad on the  shape
     comparison for \al{}-\al{}-Oxygen events is shown.}
\label{fig:costhetajac_bf}
\end{figure}

Further details on the topology of selected evaporation chains can be
obtained using the  Jacobi coordinates, suitable for 3-body
events, under the  guide of the statistical model simulated
data, where the particle emission order is known for each chain. As in our
previous paper~\cite{bib:morelli2} we restrict ourselves to the specific channel
Oxygen-2\al{} only, where the disagreement between the ex\-per\-i\-men\-tal and
predicted BR is the large; moreover, for this 3-body charged decay (possibly
perturbed by neutron emission) the use of the Jacobi coordinates is
quite well motivated.  We thus calculate the Jacobi
coordinates:
\begin{equation}
\label{eq:jacobi1}
 \epsilon =  \frac{ E_{\alpha-\alpha}} {E_{tot}}
\end{equation}
\begin{equation}
\label{eq:jacobi2}
 cos(\theta_{k})=  \vec{u}_{\rm{O}} \cdot \vec{u}_{\alpha-\alpha}
\end{equation}
where $E_{\alpha-\alpha}$ and $E_{tot}$ are the relative kinetic energy
between the \al{} pair and the total available energy, respectively.
The Jacobi angle $\theta_{k}$ is defined as the angle between the
unit vector of the relative motion of the two \al{} particles $\vec{u}_{\,\alpha-\alpha}$
and that of the Oxygen residue momentum with respect to the \al{}-\al{} center of mass.
Since there are two ways of numbering the \al{} particles,
for each event  we calculated the Jacobi coordinates
for both of them, thus forcing the cosine distribution to be symmetric
around cos($\theta_k$)=0~\cite{bib:grigorenko08}.
We study the \al{}-\al{} correlations when only one (undetected) neutron  is emitted. Thus, for the experimental data, we
limit this analysis to the
events populating the   \textit{Q}-region of Fig.~\ref{fig:hqval32a} 
between the marks  corresponding to 1n and 2n emission threshold ( that is $-20.82\,$MeV). Although not perfect,
these sharp cuts define events with  Oxygen \er{} having mass A=16.

Since we are dealing with  ``false'' 3-body events due to the additional neutron, 
the relative energy $\epsilon$ can be %strongly
  perturbed with respect to the original value:
  moreover we observed, from the MonteCarlo simulation, that the relative energy $E_{\alpha-\alpha}$ is less sensitive 
  to the neutron emission order than the angular variable. 
  Therefore, we focus only on this latter with the following remarks:
  \begin{itemize}
  \item 1st chance neutron, in the sequence \mgb{}-neutron-2\al{}; here
  the \mga{} emits 2\al{} and
  the  construction of the angular Jacobi coordinate is not perturbed
  by the neutron emission;
  \item 2nd chance neutron, in the sequence \mgb{}-\al{}-neutron-\al{}; 
  since the neutron is ejected between the 2\al{} particles,
  both vectors in eq.~\ref{eq:jacobi2} are modified and a large perturbation
  on the decay is expected.
  \item 3rd chance neutron, in the sequence \mgb{}-2\al{}-neutron; the
    neutron is emitted last from an $^{17}$O; thus
  only $\vec{u}_{\rm{O}}$ is affected by the neutron emission;
  the perturbation is low since the $^{16}$O velocity is only slightly
  affected by the recoil, due to the large mass difference between neutron and $^{16}$O.
  \end{itemize}
With this scheme in mind we can look at the experimental Jacobi angular 
distribution shown in Fig.~\ref{fig:costhetajac_bf} for the
O-2\al{} coincidences (black dots).
We see that the preferred  configuration is a rather aligned one
with the two \al{} particles reseparating close to the direction of
the recoiling \er{}. In the same picture also the prediction of
 \hf{} is drawn (red bold line); here we can
exactly choose the chain leading to $^{16}$O residues.
We see that the model  overestimates the aligned configurations. 
Before going into further detail, it is important to
check the ca\-pa\-bi\-li\-ty of \hf{} to properly reproduce the shape
of the Jacoby angular distributions. This has been done 
using the O+2\al{} events without neutron, selected as those
below the neutron threshold (right side of the mark in 
Fig.~\ref{fig:hqval32a}).  The result for this case
  is drawn in the inset of Fig.~\ref{fig:costhetajac_bf}. Within the limits of the
  statistics we observe  a noticeable agreement between experiment and model 
  that can does be used as a guide for a further
  investigation of the Jacobi coordinate for the  O+2\al{}+1n events.
  
  We start separating the cases of the three emission orders to explore
  how the cos($\theta_{k}$) distribution  changes from one to another.  
The three contributions to the total spectrum are shown in 
Fig.~\ref{fig:costhetajac_bf};  the corresponding relative weights are reported
in the left column of Table~\ref{tab:weights}. Clearly, the 2nd chance
emission  dominates while the 1st chance is a minority case.
Moreover, we see that the first chance neutron case is the only one
capable of filling the region  cos($\theta_{k}) \approx 0$ because the
corresponding shape is almost flat with a broad bump at zero. The
other two cases, instead, tend to populate  more aligned
configurations. Probably when the neutron is first-chance, it has on
average high energy and the following two (relatively slow) \al{}
produce a moderate  recoil on the heavier partner. Instead, if an \al{} particle
is emitted first, it  
has high energy (on average) and  the kick given the residue favors
polarized configurations  (cos($\theta_{k}) \approx \pm 1$).

\begin{figure}[t]
   \centering
\includegraphics[width=1\columnwidth]{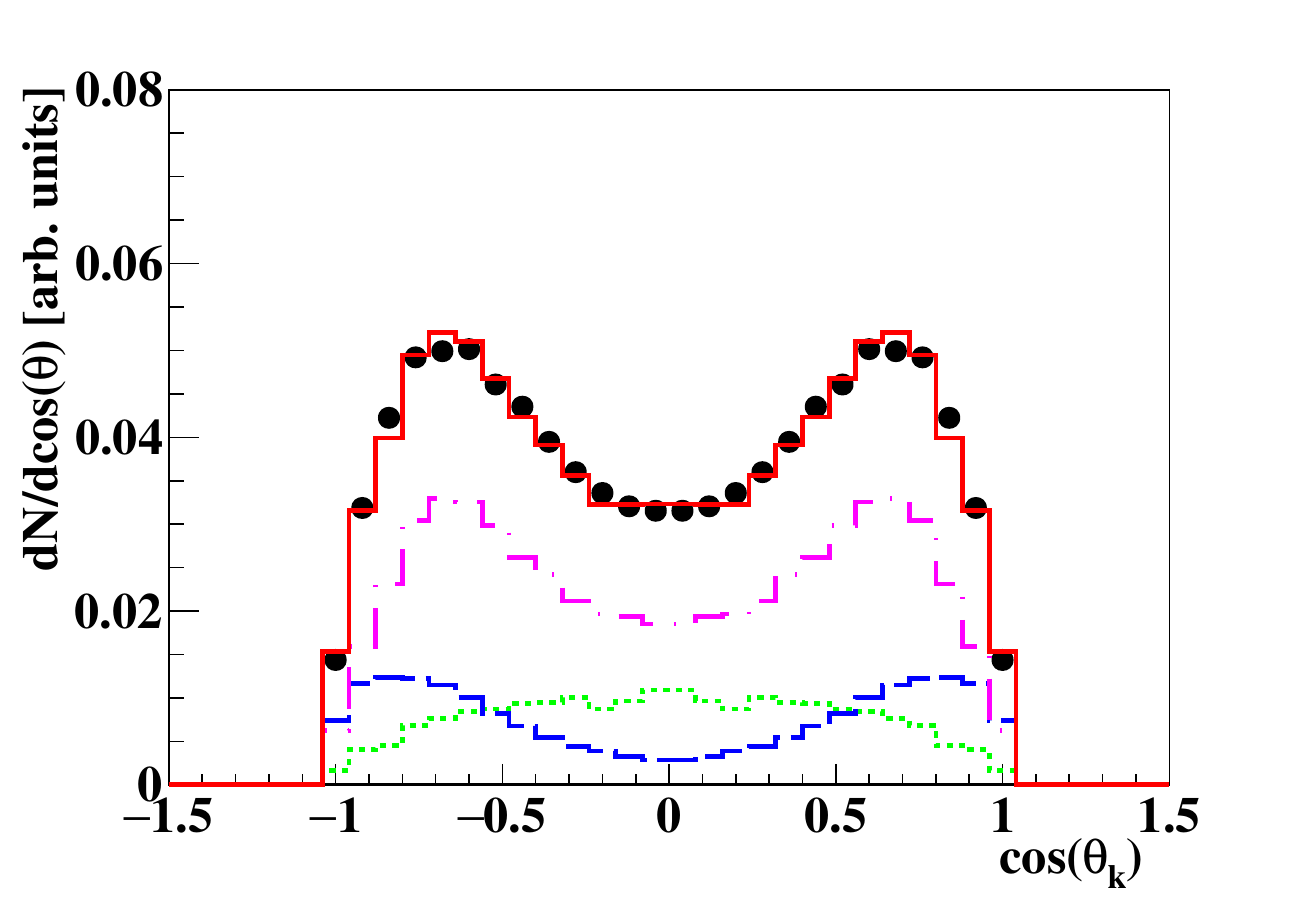}
   \caption{(Color online)
   Same as Fig.\ref{fig:costhetajac_bf} but with the weights
   assigned through the fit
   procedure explained in the text.}
\label{fig:costhetajac_af}
\end{figure}

Using the shapes predicted by the model, we estimated the new weights
of the three cases corresponding to the three emission 
orders via a fit procedure on the experimental result. Specifically, 
we looked for the minimum of a purposely defined $\chi^{2}$ variable as follows:
\begin{equation}
 \chi^{2} = \sum_{i=0}^{N} \frac{(h_{exp}(i) - h_{HF\ell}(i))^{2}}{\sigma^{2}_{exp}(i)}
\label{eq:chi2}
 \end{equation}
with
\begin{equation}
h_{HF\ell}=w_{1}\cdot h_{1st} + w_{2}\cdot{} h_{2nd} + w_{3}\cdot h_{3rd}
\label{eq:hsum}
\end{equation}
where $h_{exp}(i)$ and $h_{HF\ell}(i)$ are the values of the experimental and
simulated  spectra at the $i$-th bin, respectively;
the experimental variance for the $i$-th bin, $\sigma^{2}_{exp}(i)$,
is obtained assuming a Poisson distribution on the counts registered
in the bins;
$h_{HF\ell}$ is the total model distribution composed by the three
cases with weights
$w_{1}$, $w_{2}$ and $w_{3}$, which are the fit parameters. The
statistics of the simulated events is such that the errors on the model
distributions are negligible.
The new fitted weights are listed in the right column of
Table~\ref{tab:weights} and the high quality of the result is shown in  Fig.~\ref{fig:costhetajac_af}, where the three 
contributions 
are scaled by the new weights;  the summed curve nicely
matches with the experimental finding. 

  \begin{table}[b]
\caption{\label{tab:weights} Weights for   the 1st-2nd-3rd chance
   emission of the neutron in the evaporation chains of \mgb{} to
   O+\al{}+\al{}. The left column reports the original weights predicted by \hf{} while
   the right one
   lists the weights obtained through the  fit procedure explained in the
   text. Errors are only statistical and 
   calculated from the    $\chi^{2}$ distribution 
   of the fit procedure.} 
 \begin{ruledtabular}
 \renewcommand\arraystretch{1.1}
 \begin{tabular}{ccc}
 \bfseries  &\hf{}  & \hf{} \\
          & original code & after fit \\
  \hline
1st chance n & 5~\% & 20$\pm$2~\% \\
  \hline
2nd chance n & 70~\% & 20$\pm$2~\% \\
  \hline
3rd chance n & 25~\% & 60$\pm$4~\% \\
 \end{tabular}
 \end{ruledtabular}
 \end{table}

We can conclude that the experimental data suggest  a preferred \al{}-\al{}-n
emission as already found for the
same kind of decays from  \mga{}~\cite{bib:morelli2}. On the other
side,  the fit indicates that the neutron first chance emission is
much more probable than predicted 
by the statistical model. In these cases, after removing the neutron, the
 emission path from the decay of \mgb{} becomes almost identical to that
 of \mga{} and this situation is underestimated by \hf{} calculations. 
The fact that the evaporation chains of $^{24}$Mg and
$^{25}$Mg are similar when the excess neutron is promptly
removed  along the evaporation path could partially explain the
similarity between the decays of the two Mg nuclei and the persistence 
of the differences found between the
experimental and simulated data.

Even more interesting: we can observe  a preferential occurrence 
of chains where two \al{} are emitted one after the other. In fact, the cases
with 1st or 3rd chance n-emission are experimentally much more probable
than predicted by the \hf{} code, which instead favors  \al{}-n-\al{} chains. 
This finding 
could again hint to some \al{} cluster structure developing  
during the path to fusion, of course not include in out model.
It is very remarkable, in this direction, the message proposed in a
theoretical paper just now published~\cite{bib:schuetrumpf}. There, 
in the context of refined TDHF calculations, the authors show that \al{} clustered 
configurations occur during the pre-compound phases in  fusion reactions of 
light heavy-ions (either with $N$=$Z$ or with small neutron excess) above the barrier. 
Another interesting possibility could be the persistence, 
in nuclei at high excitation energy and with small neutron excess, 
of the linear O-\al{}-\al{} chain theoretically 
predicted in the excited spectrum of \mga{}~\cite{bib:girod, bib:ichikawa}.
These two interpretations  represent promising theoretical directions for further understanding of 
the effects presented in the this paper.

\section{\label{sec:concl}Conclusions} 

We have described the experimental results of an  experiment 
on  $^{12}$C+$^{13}$C fusion reactions at 95$\,$MeV bombarding energy, 
performed  with the apparatus \gar{}+\rc{} at the INFN Laboratori Nazionali di 
Legnaro (Italy).  
Motivated by the recent interest in the investigation of the interplay 
of nuclear structure and reaction mechanisms in  light systems and 
in continuation with our previous  studies  
~\cite{bib:morelli1, bib:morelli2},   
we focused on the decay of the hot \mgb{} compound nucleus  and  
we  studied the properties of its various decay chains.  
Specifically, the objective was to  
verify  if the disagreement of some observables with respect to refined
statistical model calculations found for the decay of the autoconjugate \mga{}
nucleus persist also with the addition of one neutron. 

Thanks to the large efficiency  and the good identification capability 
of the detectors, we could precisely select and study the various 
fusion-evaporation chains, strongly constrained by the request of 
total charge conservation. Furthermore, 
an original  attempt was also done to reconstruct the mass of the evaporation 
residues  even without measuring emitted neutrons, by exploiting the 
$Q$-value distribution for selected channels and using our  refined
Hauser-Feshbach calculations to model the various contributions.  

The main results are the following. Similarly to previous works, most 
fusion-evaporation features are well accounted for by 
a  refined version of the statistical model. 
Still, some disagreements have been found 
when looking at the details of specific evaporation chains, mainly 
those dominated by the emission of \al{} particles and reaching even-Z
\er{}.  In particular, as  
for the \mga{} case, a clear mismatch between experimental and 
predicted branching ratios  (BR) was found for the channels ending up with  
Z$_{ER}=6,8,10$, reached via pure \al{} emissions. The model strongly 
underpredicts these channels. 
 The a\-na\-lo\-gy of this result with the previous findings on  
 $^{24}$Mg~\cite{bib:morelli1} suggests that the excess neutron in \mgb{} 
 does not considerably modify   the evaporation paths,
and that possible \al{} cluster effects  still persist
in fusion reactions, not being  washed out  by the additional neutron.
This is suggested by an analysis in terms of Jacobi
the angular  coordinate, applied to the selected  decay
 O+2\al{}+n.  The deduced tendency of the \al{} particles to be  
 preferentially emitted one after the other and  
 not separated by neutron emission (as predicted  by the model) supports the argument. 

Further, even more selective experiments would be ne\-ces\-sa\-ry to better disentangle specific 
evaporation chains. 
More severe  constraints on the decay chains could be imposed by the 
coincident detection  of neutrons but this is a very 
challenging  effort presently not yet at hand. 
Alternatively, efforts can be done to improve the isotopic identification 
capability of the \er{} detectors in order to  select mass
resolved decay chains, event-by-event.
The original attempt does in this paper to deduce the \er{} masses goes in this 
direction but it is not apt to describe the mass balance for every event. 
Improvements of the isotopic separation capability of detectors are in progress
in our collaboration and the recent  developments are promising to reach even A identification for residues 
with the \rc{} telescopes, at least for light nuclei, like those studied in this 
paper. Of course, the experimental improvements should be
accompanied by more refined theoretical calculations, able to go beyond
the Hauser-Feshbach scheme and including effects related to cluster or
resonance states. The recent theoretical paper~\cite{bib:schuetrumpf} is very suggestive, showing, 
in the framework of time-dependent HF calculations, 
the formation of deformed \al{} cluster configurations  
during the path to  fusion in light heavy-ion collisions above the barrier.

\begin{acknowledgements}
Thanks are due to the accelerator staff of Legnaro Laboratories for having provided good-quality beams and to the Target
Lab of INFN-LNL and INFN-LNS for providing the targets used during this experiment.
This work was partially supported by grants from the Italian Ministry of Education, University,
and Research under Contract PRIN 2010-2011.
We also thank Giacomo Poggi for very helpful suggestion.
\end{acknowledgements}

\bibliography{biblio} 
\bibliographystyle{unsrt} 
 
\end{document}